\documentclass[journal]{IEEEtran}
\IEEEoverridecommandlockouts
\usepackage{ifpdf}
\usepackage{cite}
\usepackage{float}
\usepackage{stfloats}

\usepackage{booktabs}
\usepackage{hyperref}
\hypersetup{
	colorlinks=true,
	linkcolor=red,
	anchorcolor=red,
	citecolor=red}
\usepackage{url}

\usepackage{enumerate}
\usepackage{subfigure,graphicx,color}
\usepackage{amsthm}

\usepackage{amsmath,amssymb,amsfonts}
\usepackage{algorithmic}

\usepackage{bm}
\usepackage{bbding}

\newtheoremstyle{nonitalic}
{3pt}
{3pt}
{}
{}
{\bfseries}
{.}
{.5em}
{}

\theoremstyle{nonitalic}
\newtheorem{definition}{Definition}
\newtheorem{lemma}{Lemma}
\newtheorem{theorem}{Theorem}

\usepackage{textcomp}

\usepackage{multirow}

\usepackage{caption}
\usepackage[ruled,vlined]{algorithm2e}

\begin{document}

\title{Provably Secure Public-Key Steganography Based on Admissible Encoding}

\author{\IEEEauthorblockN{Xin Zhang, 
Kejiang Chen, 
Na Zhao, 
Weiming Zhang, 
Nenghai Yu}
\thanks{This work was supported in part by the National Natural Science Foundation of China under Grant 62472398, U2436601, U2336206 and 62402469.}
\thanks{All the authors are with University of Science and Technology of China, Hefei 230026, China, and Anhui Province Key Laboratory of Digital Security.}
\thanks{Corresponding authors: Kejiang Chen (Email:chenkj@ustc.edu.cn) and Weiming Zhang (Email:zhangwm@ustc.edu.cn)}
}
\maketitle

\IEEEdisplaynontitleabstractindextext

\IEEEpeerreviewmaketitle

\markboth{Submitted to IEEE Transactions on Information Forensics and Security}{Shell \MakeLowercase{\textit{et al.}}: Bare Demo of IEEEtran.cls for IEEE Transactions on Information Forensics and Security}

\IEEEtitleabstractindextext{%
\begin{abstract}


The technique of hiding secret messages within seemingly harmless covertext to evade examination by censors with rigorous security proofs is known as provably secure steganography (PSS). PSS evolves from symmetric key steganography to public-key steganography, functioning without the requirement of a pre-shared key and enabling the extension to multi-party covert communication and identity verification mechanisms. Recently, a public-key steganography method based on elliptic curves was proposed, which uses point compression to eliminate the algebraic structure of curve points. However, this method has strict requirements on the curve parameters and is only available on half of the points. To overcome these limitations, this paper proposes a more general elliptic curve public key steganography method based on admissible encoding. By applying the tensor square function to the known well-distributed encoding, we construct admissible encoding, which can create the pseudo-random public-key encryption function. The theoretical analysis and experimental results show that the proposed provable secure public-key steganography method can be deployed on all types of curves and utilize all points on the curve.
\end{abstract}

}

\maketitle

\IEEEdisplaynontitleabstractindextext

\IEEEpeerreviewmaketitle

\section{Introduction}
\IEEEPARstart{S}{teganography}~\cite{anderson1998limits,marvel1999spread,cox2007digital} is a covert communication method by embedding confidential data within ordinary media such as text, images, audio, and video. It protects the confidentiality of the information and conceals the presence of communication.
The essence of steganography involves the steganographer placing secret data into common media to produce stegotext, aiming for these stegotext to be indifferent from the original media. Conversely, the attacker's task~\cite {simmons1984prisoners}, referred to as steganalysis\cite{yang2018rnn},  is to identify the subtle differences between the original media and the stegotext~\cite{fridrich2012rich}.

Previous digital steganography techniques, such as least significant bit (LSB) replacement~\cite{mielikainen2006lsb}, exploiting modification direction (EMD)~\cite{zhang2006efficient} and minimum distortion steganography~\cite{6949122,
pevny2010using}, primarily focused on empirical security without theoretical validation. These methods often fail against deep learning-based steganalysis attacks~\cite{ye2017deep,boroumand2018deep}.

A natural question arises: can steganography achieve a level of security comparable to that of cryptography? To address this question, Cachin~\cite{cachin1998information} proposed the perfect security of steganography within an information-theoretic model, measured by the KL divergence $D_{\mathrm{KL}}\left(P_{\mathrm{c}}\| P_{\mathrm{s}} \right)$, which remains an ideal model unachievable in practice. Hopper \textit{et al.}~\cite{hopper2002provably} introduced a provably secure steganography based on computational complexity theory, aiming to prove that attackers with real-world capabilities cannot computationally distinguish between covertext and stegotext.

In Hopper's theory, he envisioned the concept of a perfect sampler, which has the capability of arbitrary sampling from the covertext distribution. Although this concept was not attainable then, with the development of deep learning and generative models~\cite{arjovsky2017wasserstein,jozefowicz2016exploring,prenger2019waveglow} it has now become achievable. Researchers discovered that deep generative models can serve as perfect samplers, using data generated by these models as covertext to conceal information, thus constructing provably secure steganography. Several efforts have been made to use generative models with provably secure steganography, including AC~\cite{ziegler2019neural,chen2021distribution}, ADG~\cite{zhang2021provably},  Meteor~\cite{kaptchuk2021meteor}, MEC~\cite{deperfectly} and Discop ~\cite{dingDiscopProvablySecure2023}. These works focus on the specific construction of embedding under the symmetric setting.

Recent research has increasingly focused on public-key steganography for many reasons. For instance, it does not rely on the assumption of a pre-shared key and offers significant efficiency advantages in multi-party covert communications \cite{anderson1998limits, hopper2002provably}. Furthermore, they can be extended to include identity verification mechanisms, further ensuring the security of communications. This paradigm shift marks a profound transition in steganography theory—from focusing on the design of individual methods to developing comprehensive communication protocols.

Following the conceptual framework of public-key steganography introduced by von Ahn and Hopper~\cite{von2004public}, Zhang \textit{et al.}~\cite{zhang2024provably} advanced this domain by proposing a novel public-key steganography method based on elliptic curves. Their method leverages point compression to eliminate the algebraic structure of curve points, rendering the ciphertext produced by elliptic curve public-key encryption indistinguishable from random bit strings. This method not only addresses the challenge of embedding curve points directly into steganographic covertext but also presents advantages over the integer finite field-based methods proposed by Hopper~\cite{von2004public} in terms of computational efficiency and encoding ciphertext size.

However, Zhang~\textit{et al.}'s method exhibits two limitations. First, it is constrained to work only on curves with a particular set of parameters, limiting its applicability across the broader range of elliptic curve parameters. This specificity allows attackers to potentially restrict the use of such curves. Second, the method is capable of utilizing only about half of the available curve points, necessitating the exclusion of the remaining points in practical applications. This restriction limits the deployment of algorithms and protocols that require the complete set of points, such as pairing-based methods \cite{boneh2001identity,horwitz2002toward} and deterministic cryptographic protocols like BLS \cite{boneh2003aggregate}. Consequently, this narrows the scope of public-key steganography for broader applications.

\textbf{Our method.}
Inspired by the concept of admissible encoding in elliptic curve hash schemes \cite{icart2009hash,farashahi2013indifferentiable,fouque2012indifferentiable}, we introduce a more general elliptic curve public-key steganography method based on admissible encoding. This encoding method possesses excellent properties, allowing not only for a surjection over the entire set of elliptic curve points but also enabling the derivation of a distribution indistinguishable from the uniform distribution over a finite field when its sampleable inverse function is provided. Despite its appealing attributes, finding admissible encoding that works across a broad spectrum of elliptic curves presents a significant challenge. However, by applying the tensor square function to the known well-distributed encoding, we construct admissible encoding whose properties are suitable for creating the pseudo-random public-key encryption function.

We have discovered that well-distributed encodings which can be strengthened to admissible encodings are widely present across all types of curves. Through the tensor square detailed in the main text, we confirm that our proposed method can be effectively applied to all commonly used curves.
To illustrate this, we utilize Icart encoding \cite{icart2009hash} on curves where 
$p=2\ mod\ 3$, SWU encoding \cite{brier2010efficient} on curves where $p=3\ mod\ 4$, and SW \cite{fouque2012indifferentiable} encoding on BN-like curves. We instantiate these three types of well-distributed encodings, construct their efficient sampleable inverse functions, and construct the corresponding public-key steganography schemes based on admissible encoding. Both our theoretical and experimental results prove the effectiveness and security of this scheme.

 Furthermore, in Appendix \ref{Subsec_generalization_to_hyperellipic_curves}, we list a plethora of elliptic curve well-distributed encoding methods that can be applied within our framework to construct public-key steganography.  Through the tensor exponent function, our scheme can be extended to even work on hyperelliptic curves. 

\textbf{Contributions.}
The main contributions of this paper can be summarized as follows:
\begin{itemize}
\item \textbf{Universal Applicability to All Types of Curve.} The new provable secure public-key steganography method we propose is deployable across all types of curves, significantly expanding the applicability of elliptic curve public-key steganography. 
\item \textbf{Full Utilization of Curve Points:} Our method utilizes all available points on the curve compared to the approach by Zhang~\textit{et al.}\cite{zhang2024provably}. This comprehensive utilization facilitates the implementation of complex algorithms and protocols, including pairing operations and deterministic cryptographic protocols like the BLS signature scheme, all requiring access to the full set of curve points.
\item \textbf{Efficient Instances of Commonly Used Curves.} In our instantiated schemes, we construct efficient sampleable inverse functions for the Icart, SW, and SWU methods. The corresponding public-key steganography instances operate on P-384, secp256k1, and P-256, respectively. Extensive statistical tests and steganalysis experiments validate the security of our constructions.
\end{itemize}

\section{Related Work}


\subsection{Provably Secure Steganography}
Provably secure steganography offers mathematically verifiable security to the steganography scheme, unlike its experience-based counterpart. It starts by defining a system model—symmetric or asymmetric, two-party or multi-party and then constructs a formal adversary model based on potential threats, mimicking real-world attacks. The approach uses rigorous math to reduce system security to some complex computational problems, ensuring the security is provable under assumptions.

Hopper \textit{et al.}~\cite{hopper2002provably,katzenbeisser2002defining} first introduced a framework of provably secure steganography by defining a probabilistic game named chosen plaintext attack (CPA), which models the scenario of a passive attack where the attacker hijacks the steganography encoder, which is also the working scenario for most steganalysis.

Following the definition, Hopper \textit{et al.} proposed their construction, which is based on rejection sampling using a perfect sampler defined over the channel distribution and an unbiased function. Define a channel as a distribution with timestamp: $\mathcal{C}=((c_1,t_1),(c_2,t_2),...)$, 
the perfect sampler is an oracle $\mathcal{O^\mathcal{C}}$ providing exactly the distribution of $\mathcal{C}_h$, where $h$ is noted as history. A function $f:\mathcal{C} \rightarrow R$ is called $\epsilon-biased$ if $|Pr_{x \leftarrow \mathcal{C}}[f(x)=0]-1/|R|| \leq \epsilon$. $f$ is unbiased if $\epsilon=0$.
Given hiddentext $m$, the rejection sampling is defined as follows:
$$
\text{sample}\ x \ \text{from}\ \mathcal{O^\mathcal{C}} \ \text{until}\ f(x)=m.
$$

Hopper \textit{et al.}\cite{hopper2002provably,von2004public} proved the security of their method against chosen plaintext attacks (CPA) by relying on the assumption of a perfect sampler and an unbiased function. However, the perfect sampler was not found at that time.

\subsection{Generative Model and Provably Secure Steganography}\label{Sec_PSS}
With the development of deep learning and generative
models, researchers discovered that deep generative models can serve as perfect samplers, using data generated by these models as carriers to conceal information, thus constructing provably secure steganography. 

Numerous provably secure steganography methods have been devised. Yang \textit{et al.}\cite{yang2019provably} pioneered provably secure steganography using autoregressive generative models and arithmetic coding (AC). Chen \textit{et al.}\cite{chen2021distribution} extended this to text-to-speech and text-generation tasks, respectively. Zhang \textit{et al.}\cite{zhang2021provably} introduced a method based on adaptive dynamic grouping (ADG) for provable security. Kaptchuk \textit{et al.}\cite{kaptchuk2021meteor} proposed Meteor to address randomness reuse in AC-based methods. Ding \textit{et al.}\cite{dingDiscopProvablySecure2023} presented Discop, a more efficient method based on distribution copies. de Witt \textit{et al.}\cite{deperfectly} showed that maximal transmission efficiency in perfect security equals solving a minimum entropy coupling (MEC) problem. These works emphasize the specific construction of embedding in the symmetric, two-party setting and show potential extensions to the asymmetric, multi-party setting using public-key steganography.

\subsection{Provable Secure Public-Key Steganography} \label{Def_PKS}
Recent research has shifted the focus from these symmetric
steganography methods to asymmetric methods, namely public-key steganography. This approach operates without the need for
a pre-shared key and allows expansion to multi-party covert
communication and identity verification mechanisms. 
\subsubsection{Definition}
\begin{figure}[htbp]
	\centering
	\includegraphics[width=0.45\textwidth]{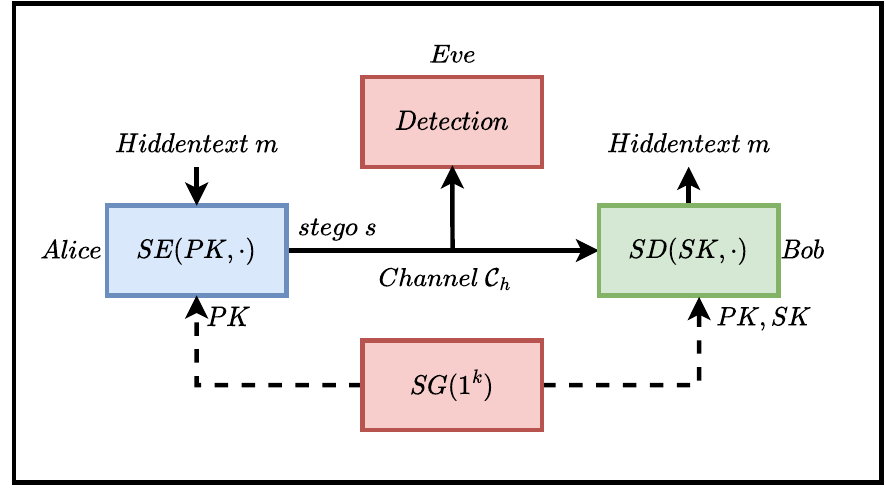}
	\caption{Diagram of the public-key steganography system.}
	\label{fig: public key}
\end{figure}

As illustrated in Fig. \ref{fig: public key}, a public-key steganography system \cite{von2004public} comprises three probabilistic algorithms $SS = (SG,SE,SD)$.  The algorithm $SG(1^k)$ generates a key pair $(PK, SK)$ from a random bitstream during the initial phase. The encoding function $SE(PK, m, h)$, using the public key $PK$, a hidden message $m$, and the history-based channel distribution $\mathcal{C}_h$, outputs stegotext $s_1, s_2, \dots, s_l$ sampled from $\mathcal{C}_h$. The decoding function $SD$ takes the secret key $SK$, a sequence of stegotext $s_1, s_2, \dots, s_l$, and the message history $h$, and returns the hidden message $m$. Both $SE$ and $SD$ have access to the channel $\mathcal{C}_h$.

\subsubsection{Existing Constructions}
Hopper \textit{et al.}\cite{hopper2005steganographic} refined the definition of the chosen plaintext attack (CPA, see Def. \ref{Def_CHA}) for public-key steganography and pointed out that constructing public-key encryption functions that output pseudorandom ciphertexts is the core to creating public-key steganography. Building on this concept, they proposed pseudorandom public-key encryption methods based on RSA and Elgamal over integer finite fields, employing the probabilistic bias removal method (PBRM) to eliminate non-random probability biases. To address the computational and encoding inefficiencies of Hopper's method, Zhang \textit{et al.}\cite{zhang2024provably} proposed an elliptic curve public-key steganography based on point compression. Specifically, they constructed a bijection from approximately half of the curve points to a uniform random string on certain curves. 

However, the previous point compression method \textit{et al.} faces two major issues: it is only applicable to a small group of curves with specific parameters, namely, \(E_{A, B}: B y^{2}=x^{3}+A x^{2}+x\ (mod\ p, p \equiv 1\ (mod\ 4), \chi\left(A^{2}-4 B\right)=-1, A B\left(A^{2}-4 B\right) \neq 0 .)\), which significantly narrows the scope of their method's applicability. Furthermore, it can only utilize about half of the curve points, forcing the exclusion of the remaining half of the points for practical deployment in public-key steganography. This limitation hinders the employment of algorithms and protocols that operate on the complete set of points, such as pairing \cite{boneh2001identity,horwitz2002toward} and BLS protocol\cite{boneh2003aggregate}, thus constraining the utility of public-key steganography in a wider array of applications.

To address the two issues presented above, we found inspiration in the elliptic curve hash scheme and introduced the potent notion of admissible encoding. Specifically, we devised a method that involves constructing a random uniform mapping from a high-dimensional finite field to an elliptic curve domain, along with a deterministic inverse mapping. These mappings are utilized to establish the framework for public-key steganography.

\subsection{Elliptic Curve Hash and Admissible Encoding}

The elliptic curve hash scheme is utilized in numerous cryptosystems that necessitate hashing an ID or something similar into an elliptic curve point. Such hash functions can substitute for any utilized within cryptosystems that rely on the random oracle model. Brier \textit{et al.} \cite{brier2010efficient} have established a sufficient condition for the construction of a hash function into an elliptic curve to be indistinguishable from a random oracle. This condition applies to hash functions of the following form:
\begin{equation}
    \mathfrak{H}(m)=F(\mathfrak{h}(m)),
\end{equation}
where  $F: S \rightarrow E\left(\mathbb{F}_{p}\right)$  is a deterministic encoding, and  $\mathfrak{h}$  is seen as a random oracle to  $S$ . Assuming that  $\mathfrak{h}$  is a random oracle, the construction is indifferentiable whenever  $F$  is an admissible encoding into  $E\left(\mathbb{F}_{p}\right)$.

\subsubsection{Admissible Encoding}
Admissible encoding is a powerful concept integral to the construction of elliptic curve hash schemes (for a detailed definition, see Def. \ref{Def_Admissible_encoding}). It has excellent attributes including \textit{computability, regularity, and samplability}, permitting a surjection across the entire set of elliptic curve points and facilitating the generation of a distribution that is indistinguishable from the uniform distribution over a higher-dimensional finite field, especially when its sampleable inverse function is provided.

We have discovered that by leveraging the properties of admissible encoding, we can effectively construct public-key encryption functions that output pseudorandom ciphertexts, thus addressing the core problem of public-key steganography. The following sections will detail this scheme and provide rigorous proof.

\subsubsection{Instantiation}
In the instantiation of our construction, 
a major technical difficulty lies in the fact that admissible encoding can hardly be constructed explicitly. To the best of our knowledge, only a special class of supersingular curves with specific parameters has an explicit expression for admissible encoding \cite{boneh2001identity}. 

Drawing on the theory by Farashahi \textit{et al.} \cite{farashahi2013indifferentiable}, by applying the tensor square function to known well-distributed encoding, we construct our admissible encoding from a two-dimensional finite field to the set of elliptic curve points. We employed various well-distributed encodings to cover commonly used curve parameters. Specifically, we utilize Icart encoding \cite{icart2009hash} on curves where \(p \equiv 2 \mod 3\), SWU encoding \cite{brier2010efficient} on curves where \(p \equiv 3 \mod 4\), and SW encoding \cite{fouque2012indifferentiable} on BN-like curves. The corresponding public-key steganography instances operate on P-384, secp256k1, and P-256, respectively. In the Appendix \ref{app: Appendix}, we also provide an extensive list of elliptic curve well-distributed encoding that can be integrated into our framework to construct public-key steganography. Through tensor square or tensor exponentiation, our scheme can be expanded to also hyperelliptic curves. Thus, we thoroughly demonstrate that our method can be applied to \textbf{all types of curves}, and due to the surjective nature of admissible encoding, our approach can utilize \textbf{all points on the curve}.

\section{Definition}
\label{sec:Def}

\begin{definition}\label{Def_Negligible Function} \textbf{(Negligible Function)} 
A function \( f:\mathbb{N} \rightarrow [0,1] \) is negligible if for any polynomial \( poly(\cdot) \), there exists a natural number \( N \in \mathbb{N} \), \(s.t.\ \forall\  n > N \), \( f(n) < \frac{1}{poly(n)} \).
\end{definition}

\begin{definition}\label{Def_Statistically_indistinguishable} \textbf{(Statistical Indistinguishable)} 
Let $X$ and $Y$ be two random variables over a set $S$. The distributions of $X$ and $Y$ are $\epsilon$-statistically indistinguishable if:
\begin{equation}
    \sum_{s \in S}|\operatorname{Pr}[X=s]-\operatorname{Pr}[Y=s]| \leq \varepsilon.
\end{equation}
\end{definition}

\begin{definition}\label{Def_provable private-key steganography}\textbf{(Basic Provably Secure Steganography Encoder)}
Let $\mathcal{C}_h$ denote the data distribution of the generative model given history $h$. Let $\mathcal{E}$ be a steganography encoder, with an output of maximum length $l$, and let $\mathcal{D}$ denote the corresponding decoding method. Assume $\mathcal{E}$ is $\epsilon$-statistically indistinguishable from the distribution of the channel $\mathcal{C}_h$, namely:

\begin{equation}\label{Eq_epsilon_encoder}
\begin{aligned}
    \mathcal{E}: m \in &\{0,1\}^{t}  \rightarrow (s_1,\cdots ,s_l) \in \mathcal{C}_h^l \\
    \sum_{c \in \mathcal{C}_h^l}|\operatorname{Pr} & [\mathcal{C}_h^l=(s_1,\cdots,s_l)] -\operatorname{Pr}[\mathcal{C}_h^l=c]| \leq \varepsilon l.
\end{aligned}
\end{equation}
The probability is calculated over uniformly distributed $t$-bit strings and accounts for all randomness in $\mathcal{E}$. As shown in \cite{hopper2002provably}, there exist constructions of provably secure basic steganography encoders, such as through rejection sampling. Additionally, as mentioned in Section \ref{Sec_PSS}, many private-key steganography methods can achieve negligible $\epsilon$-statistical indistinguishability.

\end{definition}

\begin{definition}\label{Def_ECDDH} \textbf{(Decisional Diffie-Hellman Assumption in Elliptic Curve Group)}

Let $G \triangleq E_{A,B}(\mathbb{F}_p)$ be a prime-order group of elliptic curve points, where $g$ is the generator and the order of the group is a prime $q$.
Let $\mathcal{A}$ be a probabilistic polynomial-time machine (PPTM) that takes three elements from the group \(G\) as input and outputs a single bit. The DDH advantage of $\mathcal{A}$ over the tuple \((G, g,  q)\) is defined as:

\begin{equation}
    \mathbf{A d v}_{G, g,  q}^{\mathrm{ddh}}(\mathcal{A}) \triangleq
	\left|\begin{aligned}
                &\ \underset{a, b}{\operatorname{Pr}}\left[\mathcal{A}\left(a\cdot g, b\cdot g, {ab}\cdot g\right)=1\right] \\ &-\underset{a, b, c}{\operatorname{Pr}}\left[\mathcal{A}\left(a\cdot g, b\cdot g, c\cdot g\right)=1\right]\  
            \end{aligned}
        \right|,
\end{equation}
where $a,b,c$ are chosen uniformly at random from $\mathbb{Z}_q$. 

The decisional Diffie–Hellman assumption in the Elliptic curve group is a computational hardness assumption requiring that $\mathbf{InSec}^{\mathrm{ddh}}_{G, g, q}(t) \triangleq \max_{\mathcal{A} \in \mathcal{A}(t)}\{\mathbf{A d v}_{ G,g, q}^{\mathrm{ddh}}(\mathcal{A})\}$ is negligible in $k$.

\end{definition}

\begin{definition}\label{Def_Admissible_encoding} 
\textbf{(Admissible Encoding)} \cite{brier2010efficient}
A function $F:S \rightarrow R$ between finite sets is an $\epsilon$-admissible encoding if it satisfies the following properties:
\begin{itemize}
    \item \textbf{Computability}: $F$ is computable in deterministic polynomial time.
    \item \textbf{Regularity}: For $s$ uniformly distributed in $S$, the distribution of $F(s)$ is $\epsilon$-statistically indistinguishable from the uniform distribution in $R$.
    \item \textbf{Samplability}: exists an efficient randomized algorithm $\mathcal{I}$ such that for any $r \in R$, $\mathcal{I}(r)$ induces a distribution that is $\epsilon$-statistically indistinguishable from the uniform distribution in $F^{-1}(r)$.
\end{itemize}

$F$ is an admissible encoding if $\epsilon$ is a negligible function of the security parameter.

According to the definition, namely, the regularity gives:
\begin{equation}\label{Eq_Admissible Encoding_regularity}
    \sum_{r \in R}\left|\underset{s}{\operatorname{Pr}}[F(s)=r]-\frac{1}{\# R}\right|=\sum_{r \in R}\left|\frac{\# F^{-1}(r)}{\# S}-\frac{1}{\# R}\right| \leq \varepsilon
\end{equation}
where \( \#R \), \( \#S \), and \( \#F^{-1}(r) \) correspond to the cardinalities of the sets \( R \), \( S \), and \( F^{-1}(r) \), respectively.

Intuitively, an admissible encoding is a uniform and invertible mapping from the set of preimages to the set of images. Uniform sampling over the set of preimages, when passed through the admissible encoding, results in uniform sampling over the set of images. Given an image, all corresponding preimages can be determined.

\end{definition}

\begin{definition}\label{Def_Well_distributed_encoding} 
\textbf{(Well-distributed Encoding)} \cite{farashahi2013indifferentiable}
A function $f: \mathbb{F}_{p} \rightarrow E\left(\mathbb{F}_{p}\right)$ is said to be $B$-well-distributed for some $B>0$ if, for all nontrivial characters $\chi$ of $E(\mathbb{F}_{p})$, the following bound holds:
\begin{equation}\label{equation_S_f}
    \left|S_{f}(\chi)\right| \leq B \sqrt{p}, \quad \text {where} \ S_{f}(\chi)=\sum_{u \in \mathbb{F}_{p}} \chi(f(u)).
\end{equation}
\end{definition}

\begin{definition}\label{Def_tensor_exponent fucntion}\textbf{(Tensor Exponent and Tensor Square)}
Given the curve $E\left(\mathbb{F}_{p}\right)$ over the finite field $\mathbb{F}_{p}$, the tensor exponent function $f^{\otimes s}$ is defined as follow: 
\begin{equation}
    \begin{aligned}
    f^{\otimes s}: \mathbb{F}_{p}^{s} & \rightarrow E\left(\mathbb{F}_{p}\right) \\
    (u_1, \cdots, u_s) & \mapsto f(u_1)+\cdots+f(u_s) .
    \end{aligned}
\end{equation}
Consider for a given element $D \in E\left(\mathbb{F}_{p}\right) $, the number of preimage of $f^{\otimes s}$ is defined as:

\begin{equation}
    \begin{aligned}
    N_{s}(D) & \triangleq \#\left\{\left(u_{1}, \cdots, u_{s}\right) \in\mathbb{F}_{p}^{s} \mid\right. \\
    D & \triangleq f\left(u_{1}\right)+\cdots+f\left(u_{s}\right) .
\end{aligned}
\end{equation}

Set $s$ to $2$; the function is then referred to as the tensor square function:
\begin{equation}
    \begin{aligned}
    f^{\otimes 2}: \mathbb{F}_{p}^{2} & \rightarrow E\left(\mathbb{F}_{p}\right) \\
    (u, v) & \mapsto f(u)+f(v).
    \end{aligned}
\end{equation}
\begin{equation}   
    N_{2}(D) \triangleq \#\left\{\left(u,v\right) \in\mathbb{F}_{p}^{2} \mid D \triangleq f\left(u\right)+f\left(v\right)\right\}.
\end{equation}

\end{definition}

\begin{definition}\label{IND-CPA} \textbf{(Pseudorandom Public-Key Encryption)}
Following Hopper's theory \cite{hopper2002provably}, we construct pseudorandom public-key encryption schemes that are secure in a slightly nonstandard model, denoted by IND\$-CPA, as opposed to the more standard IND-CPA. The key difference is that IND\$-CPA requires the ciphertext output by the encryption algorithm to be indistinguishable from uniformly chosen random bits, while IND-CPA only requires that an adversary cannot distinguish the encryption of two chosen plaintexts. Importantly, IND\$-CPA implies IND-CPA, but the converse does not hold, making IND\$-CPA a strictly stronger requirement. This higher standard of security is particularly suitable for applications like steganography, where indistinguishability from random noise is critical.

Consider a public-key cryptography system $CS = (G,E,D)$ and a chosen plaintext attacker $\mathcal{A}$. $\mathcal{A}$ is allowed to play a game described as follows:

\begin{itemize}
    
    \item \textbf{Key generation stage.} $(PK,SK)\leftarrow G(1^k)$.

    \item \textbf{Learning stage.} $\mathcal{A}$ sends plaintext $m_\mathcal{A}$ to the oracle and  returns $E(PK,m_\mathcal{A})$. $\mathcal{A}$ can perform this stage multiple times.
    \item \textbf{Challenge stage.} $\mathcal{A}$ sends plaintext $m \in \mathcal{M}\setminus \{m_\mathcal{A}\}$ to the oracle, which will flip a coin $b\in \{0,1\}$.
    If $b=0$, $\mathcal{A}$ obtains $c=E(PK,m)$; If $b=1$, $\mathcal{A}$ obtains $u \leftarrow U_{|E(PK,\cdot)|}$.
    \item \textbf{Guessing stage.} $\mathcal{A}$ output a bit $b^{\prime}$ as a ``guess'' about whether it obtains a plaintext or a random string.
\end{itemize}

Define the Chosen Plaintext Attack  (CPA) advantage of $\mathcal{A}$ against $S$ by:
\begin{equation}
        \mathbf{A d v}_{CS}^{\mathrm{cpa}}(\mathcal{A}, k)\triangleq 
	\left|\begin{aligned}
                &\ \underset{PK}{\operatorname{Pr}}\left[\mathcal{A}\left(PK,c\right)=1\right] \\ &-\underset{PK}{\operatorname{Pr}}\left[\mathcal{A}\left(PK,u\right)=1\right]\  
            \end{aligned}
        \right|.
\end{equation}

A public-key encryption system is indistinguishable from uniformly random bits under chosen plaintext attack (IND\$-CPA) if $\mathbf{InSec}^{\mathrm{cpa}}_{CS}(t,l,k) \triangleq \max_{\mathcal{A} \in \mathcal{A}_{(t,l)}}\{\mathbf{Adv}_{CS}^{\mathrm{cpa}}(\mathcal{A},k)\}$ is negligible in $k$.
\end{definition}

\begin{definition}\label{Def_CHA}\textbf{(Chosen Hiddentext Attack)}
Refer to Hopper \textit{et al.} \cite{hopper2005steganographic}
and Zhang \textit{et al.}'s paper \cite{zhang2024provably}, the \textbf{Threat Model} of public-key steganography is defined as follows:

Consider a public-key steganography system $SS = (SG,SE,SD)$ and an attacker $\mathcal{A}$. $\mathcal{A}$ play a game named chosen hiddentext attack (CHA) described as follows:
\begin{itemize}
    
    \item \textbf{Key generation stage.} $(PK,SK)\leftarrow SG\left(1^k\right)$.

    \item \textbf{Learning stage.} $\mathcal{A}$ sends hiddentext $m_\mathcal{A}$ and history $h_\mathcal{A}$ and gets return $SE(PK,m_\mathcal{A},h)$. $\mathcal{A}$ can perform this stage multiple times.
    \item \textbf{Challenge stage.} $\mathcal{A}$ sends hiddentext $m \in \mathcal{M}\setminus \{m_\mathcal{A}\}$ to an oracle, which will flip a coin $b\in \{0,1\}$. If $b=0$, $\mathcal{A}$ obtains $s=SE(PK,m,h)$; if $b=1$, $\mathcal{A}$ obtains $c\leftarrow \mathcal{C}_h$.
    \item \textbf{Guessing stage.} $\mathcal{A}$ outputs a bit $b^{\prime}$ as its ``guess" to determine whether it has received a stegotext or a covertext.
\end{itemize}

Define the Chosen Hiddentext Attack (CHA)\cite{hopper2002provably} advantage of $\mathcal{A}$ against $SS$ over channel $\mathcal{C}$ by:
\begin{equation}
    \mathbf{A d v}_{SS, \mathcal{C}}^{\mathrm{cha}}(\mathcal{A}, k) \triangleq
	\left|\begin{aligned}
                &\ \underset{PK}{\operatorname{Pr}}\left[\mathcal{A}\left(PK,s\right)=1\right] \\ &-\underset{PK}{\operatorname{Pr}}\left[\mathcal{A}\left(PK,c\right)=1\right]\  
            \end{aligned}
        \right|.
\end{equation}

Define the \textit{insecurity} of $SS$ over channel $\mathcal{C}$ by 
\begin{equation}
    \mathbf{InSec}^{\mathrm{cha}}_{SS, \mathcal{C}}(t,l,k)\triangleq \max_{\mathcal{A} \in \mathcal{A}_{(t,l)}}\{\mathbf{Adv}_{SS, \mathcal{C}}^{\mathrm{cha}}(\mathcal{A},k)\},
\end{equation}
where $\mathcal{A}_{(t,l)}$ is the set of all adversaries that send at most $l(k)$ bits and run in time $t(k)$. $l(k)$ and $t(k)$ are polynomials of $k$.
$SS$ is secure against CHA if $\mathbf{InSec}^{\mathrm{cha}}_{SS, \mathcal{C}}(t,l,k)$ is negligible in $k$, i.e.,  no probabilistic polynomial time (PPT) adversary can distinguish $s$ and $c$ with nonnegligible probability.

In this passive attack, the adversary $\mathcal{A}$ gains control over a steganographic encoder in the learning phase. $\mathcal{A}$ can embed specific hiddentext $m_\mathcal{A}$ into various stegotext. During the challenge phase, $\mathcal{A}$ receives a sample that might be either a stegotext generated with the encoder or a random cover. The objective in the guessing phase is for $\mathcal{A}$ to distinguish between covertext and stegotext with better accuracy than random chance. This model highlights the risks in steganography when the encoder of steganography is compromised, covering a wide range of steganalysis threats.

\end{definition}

\section{Our Proposed Method}
We will expound on our provably secure public-key steganography scheme based on admissible encoding through three sequential steps \ref{SubSec_framework}, \ref{SubSec_Proof_of_Security}, and \ref{SubSec_Instantiation}. 

Initially, we will introduce our public-key steganography framework based on admissible encoding in \ref{SubSec_framework}. The intuition is as follows: To construct a mapping from elliptic curve points to pseudorandom bitstrings, the typical approach is to first map the curve points to a finite field $\mathbb{F}_p$ and then add redundancy. If our curve parameters do not allow for mapping the curve points to a single finite field, we consider constructing an admissible encoding to map the curve points to a two-dimensional or higher-dimensional finite field.
In this section, we will detail how to construct admissible encoding from well-distributed encoding through tensor square, and then use admissible encoding to construct provably secure public-key steganography. 
Subsequently, we will present comprehensive proofs related to the aforementioned construction in \ref{SubSec_Proof_of_Security}. We will not only prove the effectiveness of the admissible encoding construction but also demonstrate that our final framework can withstand CHA attacks refer to Def. \ref{Def_CHA} in the random oracle model.
Finally, we will explain how we instantiated our framework in \ref{SubSec_Instantiation}. We adopted three different known well-distributed encodings including Icart \cite{icart2009hash}, SW \cite{shallue2006construction,fouque2012indifferentiable} and SWU \cite{ulas2007rational}, constructed efficient sampleable inverse functions for these three encodings, and deployed our public-key steganography system on curves of three types of parameters. Subsequent experiments (\ref{Sec_Experiment}) on these three instances fully demonstrated the security and effectiveness of our entire system.

\subsection{Provably Secure Public-Key Steganography Framework Based on Admissible encoding}\label{SubSec_framework}

\subsubsection{Well-distributed Encoding Strengthen to Admissible Encoding}
Since the admissible encoding can hardly be constructed explicitly. Only a special class of supersingular curves with specific parameters has an explicit expression for admissible encoding. To construct public-key steganography that is general to all types of curves, we consider the possibility of constructing admissible encoding from a two-dimensional finite field to the finite field of elliptical curves. We utilize the tensor square function to strengthen well-distributed encoding on curves with a genus of $1$ into admissible encoding (for constructions on curves not of genus $1$, see Appendix \ref{app: Appendix}).

Given a computable $B$-well-distributed encoding function \(f: \mathbb{F}_p \to E(\mathbb{F}_p)\), the admissible encoding \(F\) from the two-dimensional finite field \(\mathbb{F}_p^2\) to the finite field of elliptic curves \(E(\mathbb{F}_p)\) of genus $1$ using the tensor square function as follows:

\begin{equation}
    \begin{aligned}
    F: \mathbb{F}_{p}^{2} & \rightarrow E\left(\mathbb{F}_{p}\right) \\
    F(u, v) & = f^{\otimes 2} = f(u)+f(v).
    \end{aligned}
\end{equation}
The complete proof of the admissibility of \(F(u,v)\) can be found in Section \ref{SubSec_Proof_of_Security}, Lemma \ref{lem_Admissibility_of_tensor_square_function}.

\subsubsection{Provably Secure Public-key Steganography Framework Based on Admissible Encoding}
As illustrated in \ref{Def_PKS}, we will present a public-key steganography system consisting of three probabilistic algorithms, denoted as $SS = (SG,SE,SD)$.

Let \( k \) be the security parameter. Let \( E(\mathbb{F}_p) \) be the group of points on the elliptic curve of genus 1 defined over the finite field \(\mathbb{F}_p\), where \( p \) is a \( k \)-bit prime number. Let $g$ be a generator of the group of points on the curve, with order $q$, where $q$ is an $n$-bit prime number.
 Given $f$ a computable $B$-well-distributed encoding function \(f: \mathbb{F}_p \to E(\mathbb{F}_p)\) and suppose $\mathcal{I}$ is its sampleable inverse function, namely
\begin{equation}
\begin{aligned}
    \mathcal{I}: E(\mathbb{F}_{p}) &\rightarrow \mathbb{F}_{p}  \\
    P \rightarrow u \in  D(P) & = \left\{u \in \mathbb{F}_{p} \mid P=f\left(u\right)\right\}.
\end{aligned}
\end{equation}

Suppose $k$ is the security parameter. The public-key steganography key pair generation (SG) is defined as Alg. \ref{Alg_SG}:

\begin{algorithm}
	\renewcommand{\algorithmicrequire}{\textbf{INPUT:}}
	\renewcommand{\algorithmicensure}{\textbf{OUTPUT:}}
	\caption{\textbf{Public-key Steganography Key Pair Generation (SG)}
	                }
	\label{Alg_SG}
	\begin{algorithmic}[1]
	    \REQUIRE {$1^k \in U(|k|), (p,\  E\left(\mathbb{F}_{p}\right),\ g,\ q) $}
	    \ENSURE {$PK,\ SK$}
		\STATE Pick $x \in [0,\ q-1]$ at random;
		\STATE $PK = x \cdot g,\ SK=x$
	\end{algorithmic}  
\end{algorithm}

Define $(E_K,D_K)$ as encryption and decryption functions of a private-key encryption scheme satisfying IND\$-CPA, keyed by $\kappa$-bits key $(\kappa \le k)$. Let \( H \) be a cryptographically secure hash function \( H: \{0,1\}^k \to \{0,1\}^\kappa \). In theoretical analysis, we model \( H \) as a random oracle, an idealized function that returns an independently and uniformly distributed value for each unique input. In practice, \( H \) will be instantiated with SHA-256 or another fixed cryptographic hash function. As defined in Def.  \ref{Def_provable private-key steganography}, let $\mathcal{E}$ be a basic provably secure steganography encoder that achieves negligible $\epsilon$-statistical indistinguishability. The public-key steganography encoder (SE) is defined as Alg. \ref{Alg_SE}.

\begin{algorithm}
	\renewcommand{\algorithmicrequire}{\textbf{INPUT:}}
	\renewcommand{\algorithmicensure}{\textbf{OUTPUT:}}
	\caption{\textbf{Public-key Steganography Encoder (SE)}
	                }
	\label{Alg_SE}
	\begin{algorithmic}[1]
	    \REQUIRE {$\mathcal{E},\ m,\ (p,\  E\left(\mathbb{F}_{p}\right),\ g,\ q,\ PK),\ (f,\ \mathcal{I})$}
	    \ENSURE {$s_1,s_2,\cdots,s_*$}
            \STATE \#\# \textit{Key Deriving}
		\STATE Pick $a \in [0,\ q-1]$ at random;
		\STATE $P = a \cdot g$
            \STATE $K = H(a \cdot PK)$
            \STATE \#\# \textit{Point Hiding}
            \WHILE{True}
		\STATE Pick $v \in \mathbb{F}_{p}$ at random;
		\STATE $D \leftarrow \mathcal{I}(P-f(v))$;
            \STATE pick $i$ uniformly at random in \#$D$;
            \STATE $u \leftarrow$ $i$-th element of $D$;
            \STATE \textbf{if} $u=\varnothing$ \textbf{continue};
            \STATE \ \ \ \ \ \ \textbf{else} \textbf{break};
		\ENDWHILE
            \STATE \#\# \textit{Bias Eliminating}
            \STATE Choose $t$-bit redundancy:
            \STATE Pick $r_1 \in\left\{0, \ldots,\left\lfloor\frac{2^{k+t}-u}{p}\right\rfloor\right\}$ 
            \STATE Pick $r_2 \in\left\{0, \ldots,\left\lfloor\frac{2^{k+t}-v}{p}\right\rfloor\right\}$ 
            \STATE $(\Tilde{u},\ \Tilde{v}) = (u+r_1p,\ v+r_2p)$
            \STATE \#\# \textit{Final Encoding}
            \STATE $C_1 = (\Tilde{u},\ \Tilde{v})$
            \STATE $C_2 = E_K(m)$
            \STATE $s_1,\ s_2,\cdots,\ s_* = \mathcal{E}(C_1 || C_2)$
	\end{algorithmic}  
\end{algorithm}

The steganography encoder (SE) consists of four parts. The first part is the temporary key deriving, which samples a group element randomly using a generator of the group and multiplies its order by the receiver's public key to obtain a temporary key through the key-derived function. The second part is point hiding, which involves inverse random sampling from curve points to the two-dimensional finite field using a sampleable inverse function $f$ of well-distributed encoding. During the Point Hiding step, a random index is used to select one point from these results. If the selected point is `None', the algorithm resamples a new field element \( v \), computes \( f(v) \), and then calculates the inverse of the difference \( P - f(v) \) using the sampleable inverse function again. This ensures that a valid point is eventually chosen through random sampling. Due to the admissible encoding properties, this part's expected running time is $O(1)$ field operations, ultimately yielding two randomly uniform finite field elements, $u$, and $v$. The third part is Bias Elimination, which expands the finite field to $k+t$ bits through redundant mapping to reduce the distribution distance after truncation to binary. The parameter \( t \) in the Bias Eliminating step is set to \( \frac{k}{4} \) or \( \frac{k}{8} \) to strike a balance between security and efficiency. Larger \( t \) (e.g., \( t = \frac{k}{4} \)) leads to a smaller statistical distance between the encoded field elements and a uniform random bit string, However, increasing $t$ also adds more bits to the ciphertext, which in turn increases both storage and transmission requirements. The fourth part is the final embedding process of steganography.

The public-key steganography decoder (SD) is defined as Alg. \ref{Alg_SD}.
\begin{algorithm}
	\renewcommand{\algorithmicrequire}{\textbf{INPUT:}}
	\renewcommand{\algorithmicensure}{\textbf{OUTPUT:}}
	\caption{\textbf{Public-key Steganography Decoder (SD)}
	                }
	\label{Alg_SD}
	\begin{algorithmic}[1]
	    \REQUIRE {$\mathcal{D},\ s_{1..*},\ (p,\  E\left(\mathbb{F}_{p}\right),\ g,\ n,\ SK),\ f$}
	    \ENSURE {$m$}
            \STATE Split $C = \mathcal{D}(s_{1..*})$ into $C_1,\ C_2$;
            \STATE $(\Tilde{u},\ \Tilde{v})=C_1$
            \STATE $(u,\ v) = (\Tilde{u} \ mod\ p,\ \Tilde{v}\ mod\ p)$
            \STATE $P = f(u) + f(v)$
            \STATE $K = H(SK \cdot P)$
            \STATE $m = D_K(K, C_2)$
	\end{algorithmic}  
\end{algorithm}

The steganography decoder (SD) is straightforward. It splits the bit string extracted from the steganography, filters out the redundant parts using the module $q$, and decrypts the ciphertext by mapping the resulting finite field elements to curve points through admissible encoding.

\subsection{Proof of Security}\label{SubSec_Proof_of_Security}

\begin{figure*}
  \centering
  \includegraphics[width=\textwidth]{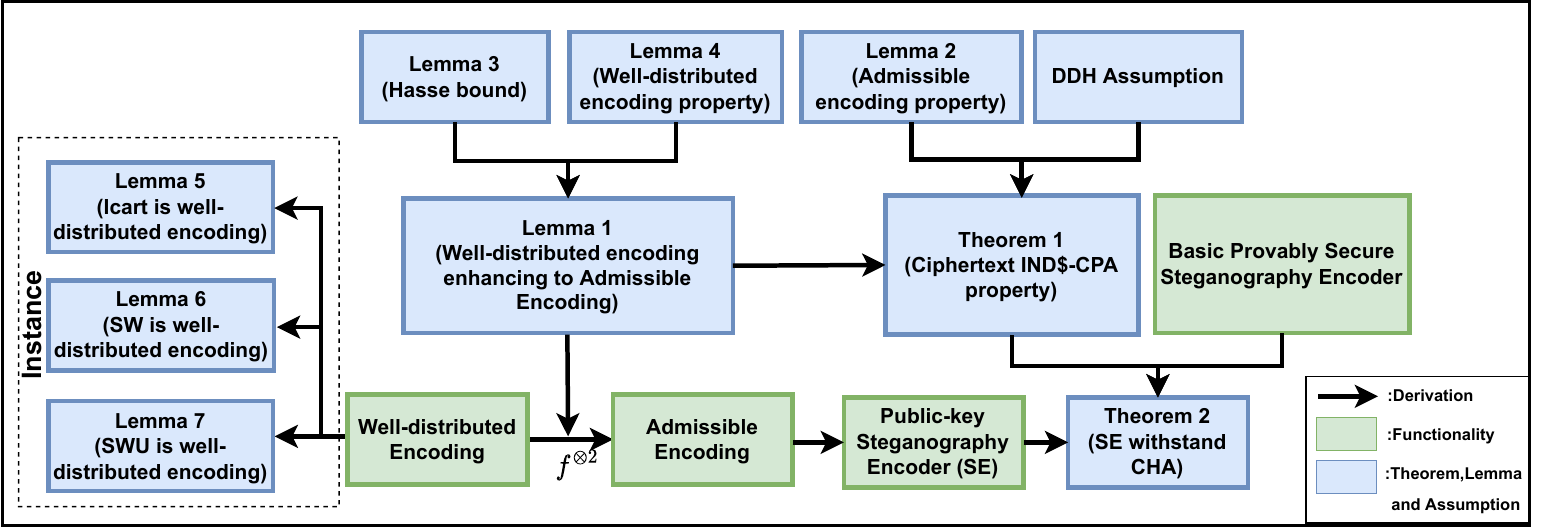}
  \caption{The theoretical framework of our public-key steganography system.}
  \label{fig:The theoretical framework of public-key steganography}
\end{figure*}

In this subsection, we will provide detailed proof of the security of our constructed public-key steganography system \(\text{SS} = (\text{SG}, \text{SE}, \text{SD})\) against CHA attacks as referred to in Def. \ref{Def_CHA}. As shown in Fig. \ref{fig:The theoretical framework of public-key steganography}, We will present the proofs of two lemmas and two theorems. Lemma \ref{lem_Admissibility_of_tensor_square_function} explains the feasibility of constructing admissible encodings from well-distributed encoding via the tensor square method. Lemma \ref{lem_admissible_encoding_properties} explains the excellent property of admissible encoding having uniform inverse sampling over the function's pre-image space. Theorem \ref{The_IND-CPA}, combined with these two lemmas and DDH assumption, establishes that the ciphertext steganography encoder used in our system has the IND\$-CPA property in the random oracle model.
Finally, we present the conclusive proof of our system's security against CHA attacks in Theorem \ref{Theorem_final_result}.

\begin{lemma}\label{lem_Admissibility_of_tensor_square_function}
Consider the tensor square function $f^{\otimes 2}$ defined as Def. \ref{Def_tensor_exponent fucntion},
if $f$ is a $B$-well-distributed encoding, and is both computable and $\epsilon^\prime$-sampleable, where $B$ is a constant and $\epsilon^\prime$ is negligible relative to the security parameter, then $f^{\otimes 2}$ is an admissible encoding from $\mathbb{F}^2_{p}$ to $E(\mathbb{F}_{p})$.

\end{lemma}
\begin{proof}

Refer to Def. \ref{Def_Admissible_encoding}, We will prove the three criteria of $f^{\otimes 2}$, namely \textit{computability, regularity, and samplability}. 
The criterion of computability is trivial for the computability of $f$.

Consider the criterion of regularity, 
 the number of preimage of $f^{\otimes 2}$ is $
    N_{2}(D) = \#\left\{\left(u,v\right) \in\left(\mathbb{F}_{p}\right)^{2} \mid D=f\left(u\right)+f\left(v\right)\right\}.
$
Since $f$ is a $B$-well-distributed encoding, according to Lemma \ref{lem_Hasse_bound} and Lemma \ref{lem_well_distributed_encoding_bound}, the statistical distance between the distribution of $F(u,v)$ for uniform $(u,v)$ and the uniform distribution on the curve can be bounded as:
\begin{equation}
    \begin{aligned}
        \sum_{D \in E\left(\mathbb{F}_{p}\right)}\left|\frac{N_{2}(D)}{p^{2}}-\frac{1}{\# E\left(\mathbb{F}_{p}\right)}\right| &\leq \frac{B^{2}}{p} \sqrt{\# E\left(\mathbb{F}_{p}\right)} \\
         \leq \frac{B^2}{p} (\sqrt{p}+1) &\leq 2B^2p^{-\frac{1}{2}},
    \end{aligned}
\end{equation}
which is a negligible function as $B$ is constant. This proves $\epsilon$-regularity.


Consider the criterion of samplability, since $f$ is $\epsilon^\prime$-sampleable, we denote $\mathcal{I}$ as its sampleable inverse function:
\begin{equation}
\begin{aligned}
    \mathcal{I}: E(\mathbb{F}_{p}) &\rightarrow \mathbb{F}_{p}  \\
    P \rightarrow u \in  D(P) & = \left\{u \in \mathbb{F}_{p} \mid P=f\left(u\right)\right\}.
\end{aligned}
\end{equation}
To show the samplability of $f^{\otimes 2}$, we construct the sampling algorithm for $f^{\otimes 2}$ as Alg. \ref{Alg_Sampling algorithm for tensor square}.

\begin{algorithm}\label{Alg_Sampling algorithm for admissible encoding}
	\renewcommand{\algorithmicrequire}{\textbf{INPUT:}}
	\renewcommand{\algorithmicensure}{\textbf{OUTPUT:}}
	\caption{\textbf{Sampling algorithm for $f^{\otimes 2}$ }
	                }
	\label{Alg_Sampling algorithm for tensor square}
	\begin{algorithmic}[1]
	    \REQUIRE {$P \in  E(\mathbb{F}_{p}), \ \mathcal{I}$}
	    \ENSURE {$(u,v) \in D^2(P) = \left\{(u,v)\in (\mathbb{F}_{p})^2 \mid P=f^{\otimes 2}\left(u,v\right)\right\}$}
		\WHILE{True}
		\STATE Pick $v \in \mathbb{F}_{p}$ at random;
		\STATE $D \leftarrow \mathcal{I}(P-f(v))$;
            \STATE pick $i$ uniformly at random in \#$D$;
            \STATE $u \leftarrow$ $i$-th element of $D$;
            \STATE \textbf{if} $u=\varnothing$ \textbf{continue};
            \STATE \ \ \ \ \ \ \textbf{else} \textbf{return} $(u,v)$;
		\ENDWHILE
	\end{algorithmic}  
\end{algorithm}
 For well-distributed encoding, the number of unmapped points is bounded. Consequently, the number of repetitions is polynomially bounded. Given that $\epsilon^{\prime}$ is negligible, the computable algorithm ensures a uniform distribution of $D^2(P)$, thereby demonstrating $\epsilon$-regularity.

\end{proof}
\begin{lemma}\label{lem_admissible_encoding_properties}
Given an $\epsilon$-admissible encoding $F:S \rightarrow R$ between finite sets and its sampleable inverse function $\mathcal{I}$, for $r$ uniformly distributed in $R$, the reversed distribution of $s=\mathcal{I}(r)$ is $2\epsilon$-statistically indistinguishable from the uniform distribution in $S$.
\end{lemma}
\begin{proof}
    Our target is to prove that for all randomness in $\mathcal{I}$, we have statistical distance between reversed distribution and uniform distribution is bounded as follows:
\begin{equation}
\begin{aligned}
    \delta:=\sum_{s \in S}\left|\underset{r}{\operatorname{Pr}}[\mathcal{I}(r)=s]-\frac{1}{\# S}\right| &\leq  2\epsilon \\
    \sum_{r \in R} \sum_{s \in F^{-1}(r)} \frac{1}{\# R}\left|\underset{r}{\operatorname{Pr}}[\mathcal{I}(r)=s]-\frac{\# R}{\# S}\right| &\leq 2\epsilon \\
    \sum_{r \in R} \sum_{s \in F^{-1}(r)}\left|\underset{r}{\operatorname{Pr}}[\mathcal{I}(r)=s]-\frac{\# R}{\# S}\right| &\leq \\ 
    \underbrace{\sum_{r \in R} \sum_{s \in F^{-1}(r)}\left|\underset{r}{\operatorname{Pr}}[\mathcal{I}(r)=s]-\frac{1}{\# F^{-1}(r)}\right|}_{\delta_1} &+ \\
    \underbrace{\sum_{r \in R} \sum_{s \in F^{-1}(r)}\left|\frac{1}{\# F^{-1}(r)}-\frac{\# R}{\# S}\right|}_{\delta_2} &\leq 2\epsilon,
\end{aligned}
\end{equation}

\begin{equation}
\begin{aligned}
    \text{Since we have}\ \delta_2 = \sum_{r \in R} \sum_{s \in F^{-1}(r)}\left|\frac{1}{\# F^{-1}(r)}-\frac{\# R}{\# S}\right|\ \ \ \ \ \ \ \  \\ = \sum_{r \in R}\left|\frac{\# F^{-1}(r)}{\# S}-\frac{1}{\# R}\right| = \sum_{r \in R}\left|\underset{s}{\operatorname{Pr}}[f(s)=r]-\frac{1}{\# R}\right|,
\end{aligned}
\end{equation}
which is the statistical distance between $F(s)$ and uniform distribution in $R$. According to the regularity of admissible encoding defined by Def. \ref{Def_Admissible_encoding}, we have $\delta_2 \le \epsilon$.

Regarding $\delta_1$, according to the samplability of admissible encoding, we have:
\begin{equation}
\begin{aligned}
    \sum_{s \in F^{-1}(r)}\left|\underset{r}{\operatorname{Pr}}
    [\mathcal{I}(r)=s]-\frac{1}{\# F^{-1}(r)}\right|&\le \epsilon \\
    \delta_1 = \sum_{r \in R} \sum_{s \in F^{-1}(r)} \frac{1}{\# R}\left|\underset{r}{\operatorname{Pr}}[\mathcal{I}(r)=s]-\frac{1}{\# F^{-1}(r)}\right| &\le \epsilon 
\end{aligned}
\end{equation}

Hence, $\delta = \delta_1+\delta_2 \le 2\epsilon$.
\end{proof}

\begin{theorem}\label{The_IND-CPA}
Let \( \mathcal{E} \) be a distribution that is \(\epsilon\)-statistically indistinguishable from the distribution of the channel \(\mathcal{C}_h\). Let \( f \) be a \( B \)-well-distributed encoding that is both computable and \(\epsilon^\prime\)-sampleable, and \( \mathcal{I} \) be a sampleable inverse function of \( f \).  Let \( H \) be a cryptographically secure hash function \( H: \{0,1\}^k \to \{0,1\}^\kappa \) which can be modeled as a random oracle. Under the decisional Diffie-Hellman (DDH) assumption in the elliptic curve group, the ciphertext \( C_1 \| C_2 \) produced by Algorithm \ref{Alg_SE} is indistinguishable from uniformly random bits under a chosen plaintext attack (IND\$-CPA security).
\end{theorem}
\begin{proof}

Define $H_0 \triangleq C_1 || C_2 = (\Tilde{u},\ \Tilde{v}) || E_{H(ax \cdot g)}(m)$, where $f(u) + f(v) = a \cdot g$.

Define $H_1$ as the variant of $H_0$ where $ax \cdot g$ is replaced by a random element of the group $E(\mathbb{F}_p)$, i.e. $H_1 \triangleq C_1 || C_2' = (\Tilde{u},\ \Tilde{v}) || E_{H(c \cdot g)}(m)$.

Define $H_2$ as the variant of $H_1$ where $H(c \cdot g)$ is replaced by a random draw from $\{0,1\}^{\kappa}$, i.e. $H_2 \triangleq C_1 || C_2'' = (\Tilde{u},\ \Tilde{v}) || E_{r \in \{0,1\}^{\kappa}}(m)$.

Define $H_3$ as the variant of $H_2$ where $C_1$ is replaced by a random draw from $\{0,1\}^{2(k+t)}$, i.e. $H_3 \triangleq \Tilde{r} || C_2'' = \Tilde{r} || E_{r \in \{0,1\}^{\kappa}}(m)$, where $\Tilde{r} \in \{0,1\}^{2(k+t)}$.

\begin{figure}[htbp]
	\centering
	\includegraphics[width=0.45\textwidth]{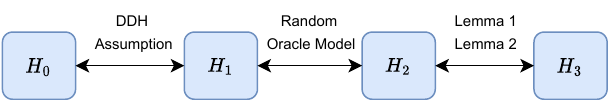}
	\caption{The hardness of distinguishing between $H_0$ and $H_3$.}
	\label{fig: fig_hardness}
\end{figure}

As shown in Fig. \ref{fig: fig_hardness}, We claim that the advantage of distinguishing between $H_0$ and $H_1$, $H_1$ and $H_2$, $H_2$ and $H_3$, as well as $H_3$ and random bits, are all negligible in $k$. 

(1) Distinguishing $H_3$ from random bits requires distinguishing $E_K(m)$ from random bits, which contradicts the IND\$-CPA security of the encryption scheme $E_K$.

(2) Distinguishing $H_2$ from $H_3$ would contradict Lemma \ref{lem_Admissibility_of_tensor_square_function} and Lemma \ref{lem_admissible_encoding_properties}, the reason is as follows:

Consider that \(C_1 = (\Tilde{u},\ \Tilde{v})\)$=(u+r_1p,\ v+r_2p)$ represents the pre-image under the admissible encoding \(F(u, v)\), where \(F(u, v) = f^{\otimes 2} = f(u) + f(v)\). This construction leverages a B-well-distributed encoding \(f\), applied through the tensor square. Accordingly, the admissibility of \(F(u, v)\) is established as per Lemma \ref{lem_Admissibility_of_tensor_square_function}, rendering \(F(u, v)\) an \(\Tilde{\epsilon}\)-admissible encoding. The value of \(\Tilde{\epsilon}\) is determined as \(\Tilde{\epsilon} = \max\{2B^2p^{-\frac{1}{2}}, {\epsilon}^{\prime}\}\), thereby affirming the admissibility criteria.

According to Lemma \ref{lem_admissible_encoding_properties}, given a sampleable inverse function of a \(\Tilde{\epsilon}\)-admissible encoding \(F\), the distribution of the pre-image is \(2\Tilde{\epsilon}\)-statistically indistinguishable from the uniform distribution over \(\mathbb{F}_{p}^{2}\). The bias-eliminating process, as detailed in Algorithm \ref{Alg_SE}, ensures that the statistical distance between the field \(\mathbb{F}_{p}\) and a \(k\)-bit uniform random string is bounded by \(2^{-t}\), evidenced by the equation:

\begin{equation}
    \sum_{u \in \mathbb{F}_{p}}\left|\frac{\left\lfloor\frac{2^{k+t}-u}{p}\right\rfloor+1}{2^{k+t}}-\frac{1}{p}\right| \leq \frac{p}{2^{k+t}} \leq 2^{-t},
\end{equation}

From this, the statistical distance between \(C_1\) and a \(2(k+t)\)-bit uniform random string is limited to \(2\cdot 2^{-t}+2\Tilde{\epsilon}\). Moreover, the statistical distance between \(C_2\) and \(l\)-bit uniform random string is \(\epsilon_0\), assuming the block cipher \(E_K\) is semantically secure. Consequently, the overall statistical distance is constrained by \(\epsilon_{c} = 2\cdot 2^{-t}+2\Tilde{\epsilon}+\epsilon_0\). For practical applications, selecting \(t\) as either \(\frac{k}{4}\) or \(\frac{k}{8}\) ensures that \(\epsilon_{c}\) is negligible in \(k\).

(3) The advantage of distinguishing $H_1$ from $H_2$ is negligible if $H$ is modeled as a random oracle $\mathcal{O}$, where the output is independently and uniformly distributed for each unique input. The reason is as follows:

In \( H_1 \), the input \( c \cdot g \) is assumed to be a uniformly distributed random element (with \( c \) being randomly generated and \( g \) being a fixed generator). Consequently, \( \mathcal{O}(c \cdot g) \) is also uniformly distributed over \( \{0,1\}^\kappa \). 

In \( H_2 \), \( H(c \cdot g) \) is directly replaced by a uniformly random value \( r \in \{0,1\}^\kappa \).

Since the distributions of \( H(c \cdot g) \) and \( r \) are identical, then \( H_1 \) and \( H_2 \) are indistinguishable in the Random Oracle Model for any polynomial-time adversary.

(4) Distinguishing $H_0$ from $H_1$ would contradict the Decisional Diffie-Hellman (DDH) assumption in the elliptic curve group as defined in Def.  \ref{Def_ECDDH}. Suppose there exists a probabilistic polynomial time algorithm $\mathcal{A}$ that can distinguish between $H_0$ and $H_1$ with non-negligible probability $\epsilon$. In that case, we can construct another probabilistic polynomial time algorithm $\mathcal{A}^{\prime}$ to break the DDH assumption. 

The construction of $\mathcal{A}^{\prime}$ is straightforward: when $\mathcal{A}^{\prime}$ receives $(a \cdot g, b \cdot g, c \cdot g)$, it sets $PK = b \cdot g$, computes $C_1 || C \triangleq (\Tilde{u}, \Tilde{v}) || E_{H(c \cdot g)}(m)$, then runs $\mathcal{A}$ on $C_1 || C$ and outputs its result. If $c = ab$, then $C_1 || C = H_0$. If $c$ is chosen uniformly at random from the group, then $C_1 || C = H_1$. 

Thus, $\mathcal{A}^{\prime}$ achieves at least $\epsilon/2$ advantage in distinguishing $(a \cdot g, b \cdot g, ab \cdot g)$ from $(a \cdot g, b \cdot g, c \cdot g)$.

\end{proof}

\begin{theorem}\label{Theorem_final_result} 
Let \( \mathcal{E} \) be a distribution that is \(\epsilon\)-statistically indistinguishable from the distribution of the channel \(\mathcal{C}_h\). Let \( f \) be a \( B \)-well-distributed encoding that is both computable and \(\epsilon^\prime\)-sampleable, and \( \mathcal{I} \) be a sampleable inverse function of \( f \).  Let \( H \) be a cryptographically secure hash function \( H: \{0,1\}^k \to \{0,1\}^\kappa \) which can be modeled as a random oracle. Under the decisional Diffie-Hellman (DDH) assumption in the elliptic curve group, the insecurity of the constructed public-key steganography system \(\text{SS} = (\text{SG}, \text{SE}, \text{SD})\) against chosen hider attacks (CHA) is negligible.
\end{theorem}
\begin{proof}
Supposed there exists a probabilistic polynomial time algorithm $\mathcal{A}$ that can distinguish between stegotext $s$ and covertext $c$ with non-negligible probability, We can construct a probabilistic polynomial time algorithm  $\mathcal{A}^{\prime}$ which plays the IND\$-CPA game: distinguishing $C_1||C_2$ from $U_{(2(k+t))}$.

The construction of $\mathcal{A}^{\prime}$ is straightforward:
$\mathcal{A}^{\prime}$ first chooses history $h_\mathcal{A}$ and a message $m_\mathcal{A}$ and then runs $\mathcal{A}$ to go through the key generation stage.
During the Challenge stage, $\mathcal{A}^{\prime}$  picks plaintext $m\in \mathcal{M} \setminus \{m_\mathcal{A}\}$ and sends it to the oracle. The oracle will flip a coin $b$, where for $b=0$, $\mathcal{A}^{\prime}$  obtains $C_1||C_2$, and for $b=1$, $\mathcal{A}^{\prime}$  obtains $u \leftarrow U_{(2(k+t))}$.
After receiving the oracle's return, $\mathcal{A}^{\prime}$ encodes it into multimedia data using the generated model and sends it to $\mathcal{A}$ to make a guess about the coin flip. 
$\mathcal{A}$ outputs a bit $b^{\prime}$ as its answer, which is also $\mathcal{A}^{\prime}$'s answer. The total time of the whole process is  $t+O(lk)$.

If $b = 0$, then $s \leftarrow \mathcal{E}(PK, m, h)$, so $\Pr[\mathcal{A}^{\prime}(PK, C_1 || C_2) = 1] = \Pr[\mathcal{A}(PK, s) = 1]$. 
If $b = 1$, then $c \leftarrow U_{(2(k+t))}$, so $s$ is distributed identically to $\mathcal{C}_h^l$. Thus, $|\Pr[\mathcal{A}^{\prime}(PK, u) = 1] - \Pr[\mathcal{A}(PK, \mathcal{C}_h^l) = 1]| \leq \epsilon l$ 
because $\mathcal{E}$ is $\epsilon$-statistically indistinguishable from the distribution of the channel $\mathcal{C}_h^l$.

Combining the cases, we have 
\begin{equation}
\begin{aligned}
    &|\Pr[\mathcal{A}(PK, s)-\Pr[\mathcal{A}(PK, \mathcal{C}_h^l) = 1]| \\ &\qquad =|\Pr[\mathcal{A}^{\prime}(PK, C_1||C_2)- \Pr[\mathcal{A}(PK, \mathcal{C}_h^l) = 1]|
    \\ &\qquad \le |\Pr[\mathcal{A}^{\prime}(PK, C_1||C_2)-\Pr[\mathcal{A}^{\prime}(PK, u) = 1]|\\ &\qquad \qquad + |\Pr[\mathcal{A}^{\prime}(PK, u) = 1]-\Pr[\mathcal{A}(PK, \mathcal{C}_h^l) = 1]| \\ &\qquad \le \mathbf{A d v}_{CS}^{\mathrm{cpa}}(\mathcal{A}, k)+ \epsilon l, \\ &i.e. \quad \mathbf{A d v}_{SS, C}^{\mathrm{cha}}(\mathcal{A}, k)  \le \mathbf{A d v}_{CS}^{\mathrm{cpa}}(\mathcal{A}^{\prime}, k)+ \epsilon l.
\end{aligned}
\end{equation}

Thus, if $\mathbf{Adv}_{SS, \mathcal{C}}^{\mathrm{cha}}(\mathcal{A}, k)$ is non-negligible, then $\mathbf{Adv}_{CS}^{\mathrm{cpa}}(\mathcal{A}^{\prime}, k) \ge \mathbf{Adv}_{SS, \mathcal{C}}^{\mathrm{cha}}(\mathcal{A}, k) - \epsilon l$ is also non-negligible, which contradicts IND\$-CPA property proved in Theorem \ref{The_IND-CPA}.

Hence, we have comprehensively completed the proof.

\end{proof}

\subsection{Instance}\label{SubSec_Instantiation}
To demonstrate the generality of our framework and provide instances for the practical deployment of public-key steganography, we will explain how we instantiated our framework on commonly used curves. We adopted three different known well-distributed encodings including Icart \cite{icart2009hash}, SW \cite{shallue2006construction,fouque2012indifferentiable} and SWU \cite{ulas2007rational}, constructed efficient sampleable inverse functions for these three encodings, and deployed our public-key steganography system on curves of three types of parameters, namely P-384, secp256k1 and P-256.

\textbf{Icart's encoding.} 
Icart \textit{et al.}\cite{icart2009hash} proposed an encoding method that utilizes the cube root of the curve equation, employing radicals whose degrees are prime relative to the order of the multiplicative group. Consider the curve $E_{a,b}: y^2 = x^3 + ax + b$ over the field $E_{a,b}\left(\mathbb{F}_{p}\right)$ where $p>3$ and $p =2\ (mod\ 3)$.
In these finite fields, the function $x \mapsto x^{3}$ is a bijection with inverse function $x \mapsto x^{1 / 3}=x^{\left(2 p-1\right) / 3}$. The Icart's encoding is defined as follows:

\begin{equation}\label{equation_Icart_encoding}
    \begin{aligned}
    f: \mathbb{F}_{p} & \mapsto E_{a,b}\left(\mathbb{F}_{p}\right) \\
    u & \mapsto(x, y) \\
    \text{where}\ x&=\left(v^{2}-b-\frac{u^{6}}{27}\right)^{1 / 3}+\frac{u^{2}}{3}, \\
    y&=u x+v, \\
    v&=\frac{3 a-u^{4}}{6 u} .
    \end{aligned}
\end{equation}

The Icart's encoding is a ($12+3p^{-\frac{1}{2}}$)-well-distributed encoding according to lemma. \ref{lem_Icart_bound}. To construct its sampleable inverse function $\mathcal{I}$,  it is necessary to solve the quartic equation over a finite field:

\begin{equation}\label{Eq. quartic equation}
    u^{4}-6 u^{2} x+6 u y-3 a=0 \text {. }
\end{equation}

We use the Berlekamp algorithm \cite{shoup1995new} to get the solution set, and $\mathcal{I}(P,i)$ return $i$-th root of the Eq. \ref{Eq. quartic equation}. Solving polynomial equations of degree \( d \) over a finite field can be achieved in \( O(d^3 +M(d)\cdot \log d\cdot \log p) \) scalar operations, where $M(d) = d\cdot \log d \cdot \log \log d$. In the case of a degree-4 polynomial (\( d = 4 \)) and a \( k \)-bit prime \( p \), this complexity becomes \(  O(k) \) scalar operations. Since scalar operations such as division, inversion, squareness check, and square root can be implemented in \( O(k^3) \) within the finite field \( \mathbb{F}_p \) assuming that multiplication is implemented in \( O(k^2) \),  the overall time complexity of the algorithm is \( O(k^4) \).

For our entire public-key steganography system using Icart's encoding, we have chosen the P-384 elliptic curve. This curve is part of the NIST (National Institute of Standards and Technology) suite of standards for elliptic curve cryptography, renowned for its strong security properties and efficiency in cryptographic operations. The choice of P-384 specifically offers a good balance between computational efficiency and security, making it well-suited for the demanding requirements of public-key steganography. The parameter specifications of the P-384 curve are as follows:
\begin{equation}
\text{P-384:}
\begin{cases}
    p = 2^{384} - 2^{128} - 2^{96} + 2^{32} - 1, \\
    y^2 = x^3 - 3x + b.
\end{cases}
\end{equation}

\textbf{SW encoding.} Shallue and van de Woestijne \cite{shallue2006construction} construct an encoding function $f:\mathbb{F}_{p} \rightarrow E\left(\mathbb{F}_{p}\right)$
based on the construction of explicit rational curves on a surface associated with the target curve. It is worth emphasizing that this encoding can be used in any curves that can be expressed in the Weierstrass form: $E: y^{2}=g(x)=x^{3}+A x^{2}+B x+C$. Here for simplicity, we consider SW encoding applied to a BN curve \cite{barreto2005pairing} with $E: y^{2}=g(x)=x^{3}+b$. The SW encoding is defined as follows:

\begin{equation}\label{equation_SW_encoding}
    \begin{aligned}
    f: \mathbb{F}_{p} & \mapsto E\left(\mathbb{F}_{p}\right) \\
    t &\mapsto\left(x_{i}, \chi_{p}(t) \cdot \sqrt{g\left(x_{i}\right)}\right) \\
    x_{1}(t)&=\frac{-1+\sqrt{-3}}{2}-\frac{\sqrt{-3} \cdot t^{2}}{1+b+t^{2}}, \\
    x_{2}(t)&=\frac{-1-\sqrt{-3}}{2}+\frac{\sqrt{-3} \cdot t^{2}}{1+b+t^{2}}, \\
    x_{3}(t)&=1-\frac{\left(1+b+t^{2}\right)^{2}}{3 t^{2}}.
    \end{aligned}
\end{equation}
where for each $t$, $i \in \{1,2,3\}$ is the smallest index such that $g(x_i)$ is a square in $\mathbb{F}_{p}$.

The SW encoding is a ($62+O(p^{-\frac{1}{2}})$)-well-distributed encoding according to lemma. \ref{lem_SW_bound}. To construct its sampleable inverse function $\mathcal{I}$,  we need to solve for \( t \) when given a curve point \((x, y)\). Accordingly, we present Alg. \ref{Alg_Inverse function for SW encoding} as follows:

\begin{algorithm}\label{Alg_Inverse function for SW encoding}
	\renewcommand{\algorithmicrequire}{\textbf{INPUT:}}
	\renewcommand{\algorithmicensure}{\textbf{OUTPUT:}}
	\caption{\textbf{Sampleable Inverse Function $\mathcal{I}$ for SW Encoding}
	                }
	\begin{algorithmic}[1]\label{Alg_Inverse function for SW encoding}
	    \REQUIRE {$(x,\ y)=P \in  E(\mathbb{F}_{p}): y^2=x^3+b \ (mod\ p)$}
	    \ENSURE {$D(P) = \left\{u \in \mathbb{F}_{p} \mid P=f\left(u\right)\right\}$}
		\STATE $c_1 = \sqrt{-3},\ c_2=(c_1-1)/2,\ c_3 = (-c_1-1)/2$
            \STATE $z = 2x+1$
            \STATE $s_1 = (1+b)(c_1-z)/(c_1+z)$
            \STATE $s_2 = (1+b)(c_1+z)/(c_1-z)$
            \STATE $s_3 = (z+\sqrt{(z^2-16(b+1)^2)})/4$
            \STATE $s_4 = (z-\sqrt{(z^2-16(b+1)^2)})/4$
            \STATE \textbf{If} $c_2-\frac{c_1s_1}{1+b+s_1}$ is square \textbf{then} $s_2 = s_3 = s_4 = \perp$;
            \STATE \textbf{If} $c_3+\frac{c_1s_2}{1+b+s_2}$ is square \textbf{then} $s_3 = s_4 = \perp$;
            \STATE $u_1,u_2,u_3,u_4 = \sqrt{s_1},\sqrt{s_2},\sqrt{s_3},\sqrt{s_4}$ \STATE set $u_i = -u_i$ if $is\_odd(u_i)\neq is\_odd(y)$ ,$\forall i \in 1..4$
            \RETURN {$D(P)=\{u_1,u_2,u_3,u_4\}$}
	\end{algorithmic}  
\end{algorithm}

In Alg. \ref{Alg_Inverse function for SW encoding}, we search for a feasible solution for \( t^2 \) using the formulas for \( x_1(t) \), \( x_2(t) \), and \( x_3(t) \) one by one. For each formula, we check whether \( t^2 \) is a square in \( \mathbb{F}_{q} \). Since \( i \in \{1,2,3\} \) is the smallest index such that \( g(x_i) \) is a square in \( \mathbb{F}_{q} \), we only take the first \( t^2 \) that satisfies this condition as our solution. Therefore, the correctness of the algorithm is established.

As for complexity, the algorithm we provided performs a series of calculations and checks to determine whether \( t^2 \) is a square in \( \mathbb{F}_p \) for each of the three possible cases: \( x_1(t) \), \( x_2(t) \), and \( x_3(t) \). Since each step (including addition, subtraction, multiplication, inversion, square root calculation, and squareness checking) requires a constant number of field operations, the algorithm's complexity is primarily determined by the most computationally expensive operations: square root calculation and squareness checking. The overall time complexity of the algorithm is \( O(k^3) \).

For our entire public-key steganography system using SW encoding, we have chosen curves with characteristics similar to BN curves, including widely used ones like secp256k1, due to their potential suitability for pairing deployment. The parameter specifications of the P-384 curve are as follows:
\begin{equation}
\text{secp256k1:}
\begin{cases}
    p = 2^{256} - 2^{32} - 997, \\
    y^2 = x^3 + 7.
\end{cases}
\end{equation}

\textbf{SWU encoding.} Ulas \cite{ulas2007rational} enhanced the SW encoding to diminish its complexity for curves defined by the equation $E:y^2 = g(x)=x^3+ax+b$ where $a,b \ne 0$ and $p = 3 (\mod 4)$. The SWU encoding is defined as follows:

\begin{equation}\label{equation_SWU_encoding}
    \begin{aligned}
    f: \mathbb{F}_{p} & \mapsto E\left(\mathbb{F}_{p}\right) \\
    t &\mapsto\left(x_{i}, \chi_{p}(t) \cdot \sqrt{g\left(x_{i}\right)}\right) \\
    x_1(t)&=\frac{-b}{a}\left(1+\frac{1}{t^{4}-t^{2}}\right), \\
    x_2(t)&=\frac{bt^2}{a}\left(1+\frac{1}{t^{4}-t^{2}}\right). \\
    \end{aligned}
\end{equation}
where for each $t$, $i \in \{1,2\}$ is the smallest index such that $g(x_i)$ is a square in $\mathbb{F}_{p}$.

The SWU encoding is a $(52+151p^{-\frac{1}{2}})$-well-distributed encoding according to lemma. \ref{lem_SWu_bound}. To construct its sampleable inverse function $\mathcal{I}$, 
we need to solve for \( t \) when given a curve point \((x, y)\). Accordingly, we present Alg. \ref{Alg_Inverse function for SWU encoding} as follows:

\begin{algorithm}\label{Alg_Inverse function for SWU encoding}
	\renewcommand{\algorithmicrequire}{\textbf{INPUT:}}
	\renewcommand{\algorithmicensure}{\textbf{OUTPUT:}}
	\caption{\textbf{Sampleable Inverse Function $\mathcal{I}$ for SWU Encoding}
	                }
	\begin{algorithmic}[1]\label{Alg_Inverse function for SWU encoding}
	    \REQUIRE {$(x,\ y)=P \in  E(\mathbb{F}_{p}): y^2=x^3+b \ (mod\ p)$}
	    \ENSURE {$D(P) = \left\{u \in \mathbb{F}_{p} \mid P=f\left(u\right)\right\}$}
		\STATE $\delta_1 = 1-\frac{4b}{ax+b}$
            \STATE $\delta_2 = (\frac{ax}{b}+1)^2-4(\frac{ax}{b}+1)$
            \STATE $s_1 = \frac{1-\sqrt{\delta_1}}{2},\ s_2 = \frac{1+\sqrt{\delta_1}}{2}$
            \STATE $s_3 = (\frac{ax}{b}+1-\sqrt{\delta_1})/{2},\ s_4 = (\frac{ax}{b}+1+\sqrt{\delta_1})/{2}$
            \STATE \textbf{If} $\frac{-b}{a}(1+\frac{1}{(s_1)^2-s_1})$ is square or $\frac{-b}{a}(1+\frac{1}{(s_2)^2-s_2})$ is square \textbf{then} $s_3 = s_4 = \perp$;
            \STATE $u_1,u_2,u_3,u_4 = \sqrt{s_1},\sqrt{s_2},\sqrt{s_3},\sqrt{s_4}$ \STATE set $u_i = -u_i$ if $is\_odd(u_i)\neq is\_odd(y)$, $\forall i \in 1..4$
            \RETURN {$D(P)=\{u_1,u_2,u_3,u_4\}$}
	\end{algorithmic}  
\end{algorithm}

In Alg. \ref{Alg_Inverse function for SWU encoding}, we search for a feasible solution for \( t^2 \) using the formulas for \( x_1(t) \) and \( x_2(t) \). For each formula, we check whether \( t^2 \) is a square in \( \mathbb{F}_{p} \). Since \( i \in \{1,2\} \) is the smallest index such that \( g(x_i) \) is a square in \( \mathbb{F}_{p} \), we only take the first \( t^2 \) that satisfies this condition as our solution. Therefore, the correctness of the algorithm is established.

As for complexity, since each step (including addition, subtraction, multiplication, inversion, square root calculation, and squareness checking) requires a constant number of field operations, the algorithm's complexity is primarily determined by the most computationally expensive operations: square root calculation and squareness checking. The overall time complexity of the algorithm is \( O(k^3) \).

For our entire public-key steganography system using SWU encoding, we have chosen the P-256 elliptic curve. This curve is part of the NIST suite of standardized curves. The parameter specifications of the P-256 curve are as follows:
\begin{equation}
\text{P-256:}
\begin{cases}
    p = 2^{256} - 2^{224} + 2^{192} + 2^{96} - 1, \\
    y^2 = x^3 - 3x + b.
\end{cases}
\end{equation}

Comparing the algorithmic complexities of these three methods, SWU may offer better efficiency in terms of runtime due to its fewer field operations and deployment on smaller finite fields, and Icart generally requires the most field operations, which could lead to relatively the slowest performance. However, the key strength of these three methods lies in their applicability to different types of elliptic curves, ensuring that our approach can be extended to a wide range of curve parameters. This flexibility underscores the robustness and versatility of our methods in diverse elliptic curve settings.

Up to now, we have instantiated our public-key steganography framework on three categories of commonly used curves, employing three distinct methods of well-distributed encoding to develop admissible encodings, which has effectively demonstrated the versatility and efficiency of our proposed framework.
\section{Experiments}\label{Sec_Experiment}
In this section, we evaluate the pseudorandomness of the proposed elliptic curve pseudorandom public-key encryption algorithm through statistical tests. Additionally, we validate the security of our proposed public-key steganography instance \footnote{Our code can be found on \href{https://github.com/XinZhang1999/Public-key-Discop}{https://github.com/XinZhang1999/Public-key-Discop}.} through steganalysis experiments.

\subsection{Statistical Test for Pseudorandomness}
To evaluate the pseudorandomness of our elliptic curve pseudorandom public-key encryption algorithm, we utilized the NIST SP 800-22 test suite.  \footnote{You can download the NIST SP 800-22 test suite directly from \href{https://csrc.nist.gov/projects/random-bit-generation/documentation-and-software}{https://csrc.nist.gov/projects/random-bit-generation/documentation-and-software}.}

A key pair was generated by Alg. \ref{Alg_SG}, and we generated the ciphertext $C_1 || C_2$ by Alg. \ref{Alg_SE}. This process is repeated to compile a binary string exceeding \(10^8\) bits, then segmented into 100 equal-length streams for 15 statistical tests. This procedure was replicated with various key pairs, yielding consistent results. Below in Tab. \ref{tab:combined_test}, we detail the outcomes from a representative trial:

\begin{table}[h]
	\centering
	\caption{Statistical tests for pseudorandomness of ciphertext $C_1||C_2$ in Alg.\ref{Alg_SE} using Icart's, SW, and SWU methods.}
	\label{tab:combined_test}
	\begin{tabular}{@{}cccccc@{}}
	\toprule
	\multirow{2}{*}{Statistical Test} & \multicolumn{3}{c}{P-VALUE} & \multicolumn{1}{c}{Result} \\ \cmidrule(l){2-6} 
	 & Icart's & SW & SWU &  Combined \\ \midrule
	Frequency & $0.924$ & $0.514$ & $0.181$ &  PASS \\
	BlockFrequency & $0.719$ & $0.595$ & $0.911$ &  PASS \\
	CumulativeSums & $0.739$ & $0.003$ & $0.058$ &  PASS \\
	Runs & $0.224$ & $0.437$ & $0.911$  & PASS \\
	LongestRun & $0.455$ & $0.955$ & $0.008$ & PASS &   \\
	Rank & $0.129$ & $0.289$ & $0.834$  & PASS \\
	FFT & $0.045$ & $0.080$ & $0.021$  & PASS \\
	NonOverlappingTemplate & $0.224$ & $0.058$ & $ 0.616$  & PASS \\
	OverlappingTemplate & $0.867$ & $0.455$ & $0.129$  & PASS \\
	Universal & $0.699$ & $0.924$ & $0.383$  & PASS \\
	ApproximateEntropy & $0.236$ & $0.616$ & $0.383$ & PASS \\
	RandomExcursions & $0.037$ & $0.324$ & $0.500$  & PASS \\
	Serial & $0.474$ & $0.946$ & $0.851$  & PASS \\
	LinearComplexity & $0.171$ & $0.657$ & $0.436$  & PASS \\ \bottomrule
\end{tabular}
\end{table}

To assess the randomness of the encrypted data, we evaluated the proportion of sequences that passed a specific statistical test. Using a significance level of $\alpha=0.01$ and considering $n=100$ sequences, we determined the acceptable range of proportions using the confidence interval formula $\hat{p} \pm 3\sqrt{\frac{\hat{p}(1-\hat{p})}{n}}$, where $\hat{p}=1-\alpha$. If the proportion falls outside of this interval, it suggests evidence of nonrandomness.

For $n=100$ and $\alpha=0.01$, the calculated confidence interval is $0.99 \pm 3\sqrt{\frac{\hat{p}(1-\hat{p})}{n}} = 0.99 \pm 0.0298$ (i.e., the proportion should be greater than 0.9602).

Based on Table \ref{tab:combined_test}, the ciphertext generated by our designed public-key steganography encoder \(SE\) based on three methods have all successfully passed 15 types of tests in the NIST SP 800-22 suite. This ensures that the stegotext and covertext produced through this ciphertext steganography are indistinguishable, confirming the effectiveness of our approach in maintaining the indistinguishability between stegotext and covertext and the universality of our framework.

\subsection{Steganalysis Experiments}

Although we have proven the security of the public-key steganography framework in Theorem \ref{Theorem_final_result}, we continue to engage in steganalysis to differentiate between covers (generated by random sampling) and stegotext (generated by steganographic sampling), ensuring the integrity of this research. Steganalysis is a technology used to discern stegotext from covertext, primarily relying on binary classifiers:
\begin{equation}
        \mathbf{F}( {X})=\left\{
        \begin{array}{rlc}
            0, & \text{ if }\Phi( {X})<0.5\\
            1, & \text{ if }\Phi( {X})\ge 0.5,
        \end{array}
    \right.
\end{equation}
Where $\Phi({X}) \in [0, 1]$ represents the probability that the input $ {X}$ is covertext ($\mathbf{F}=0$) or stegotext ($\mathbf{F}=1$). A false alarm occurs when $ {X}$ is a covertext while $\mathbf{F}=1$, and a missed detection occurs when $ {X}$ is a stegotext while $\mathbf{F}=0$. False alarm and missed detection are defined as follows, respectively:

\begin{align}
P_\text{FA} &= \Pr\{\mathbf{F}( {X})=1\mid {X}\in \mathcal{C}\}, \\
P_\text{MD} &= \Pr\{\mathbf{F}( {X})=0\mid {X}\in \mathcal{S}\}.
\end{align}

Here, $\mathcal{C}$ and $\mathcal{S}$ represent the covertext set and the stegotext set, respectively. Then, the overall performance is determined by the probability of detection error computed from $P_\text{FA}$ and $P_\text{MD}$ as follows:

\begin{equation}
P_\text{E} = \frac{P_\text{FA} + P_\text{MD}}{2}.
\end{equation}

We instantiated our public-key steganography framework based on admissible encoding using Discop \cite{dingDiscopProvablySecure2023} with Llama-2-7B on P-384, secp256k1, and P-256, respectively. Covertext and stegotext were generated in pairs with identical contexts, and all covertext and stegotext were generated with a truncation parameter of $p = 0.95$. To establish the dataset, we randomly selected various short sentences as contexts and generated 10,000 covers and 10,000 stegotext. The example of covertext and stegotext pairs are shown in Tab. \ref{tab:Example_of_cover_stego_pairs}. We then employed three linguistic steganalyzers, including FCN~\cite{yang2019fast}, R-BiLSTM-C~\cite{niuHybridRBILSTMCNeural2019}, and BiLSTM-Dense~\cite{yangLinguisticSteganalysisDensely2020}.
The steganalysis experiments were conducted on a dataset of 10,000 samples of stegotext and covertext, divided into training, validation, and test sets in ratios of 3:1:1. The result of steganalysis for our provably secure public-key steganography based on admissible encoding is shown in Tab. \ref{tab:Steganalysis}.

\begin{table}[tb]
\scriptsize
\centering
\caption{An example of covertext and stegotext pairs under truncation parameter $p=0.95$.}
\label{tab:Example_of_cover_stego_pairs}
\begin{tabular}{l}
\toprule
\begin{tabular}[c]{p{0.9\columnwidth}}
\textbf{Context:}\\ Years later, he would find himself\end{tabular} \\
\midrule
\begin{tabular}[c]{p{0.9\columnwidth}}
\textbf{Covertext:}\\ Years later, he would find himself standing in front of the same old oak tree, but this time it was different.
The once-thriving tree had withered and died, its branches bare and lifeless. The man felt a pang of sadness as he gazed upon the remnants of his childhood memories. He remembered the countless hours he spent playing under that very tree, chasing after acorns and watching them fall to the ground.
But now, everything seemed so different. The world had moved on without him, leaving behind only shadows of what once was. The man couldn't help but feel lost and alone, like a ghost from a bygone era.
As he stood there, mired in thought, a gentle breeze rustled through the leaves, causing them to whisper secrets of the past. The man closed his eyes, letting the memories wash over him like ...
\end{tabular} \\
\midrule
\begin{tabular}[c]{p{0.9\columnwidth}}
\textbf{Stegotext:}\\ Years later, he would find himself standing in front of the same building, but it was no longer a place of wonder and curiosity. It had been transformed into a soulless office tower, devoid of any character or charm. He couldn't help but feel a sense of loss as he gazed upon its sterile facade, remembering the countless hours he had spent exploring every corner of that magical place.
As he walked away from the building, he realized that his own life had taken on a similar tone. Once full of excitement and possibility, it now felt stifling and predictable. He found himself longing for the freedom and adventure of those early days, when anything seemed possible.
But even as he looked back wistfully at the past, he knew that he couldn't turn back the clock. Life had moved forward, and with it, so had he. The world had changed, and he had changed along ...
\end{tabular} \\
\bottomrule\\
\end{tabular}
\end{table}

\begin{table}[h]
    \centering
    \caption{Steganalysis results for three public-key steganography instances.}
    \label{tab:Steganalysis}
    \begin{tabular}{cccccc}
    \toprule
    & Steganalyzer & $P_\text{E}$ & $P_\text{FA}$ & $P_\text{MD}$ \\
    \midrule
    \multirow{4}{*}{Instance on P-384}   
     & FCN & 49.68\%  & 49.70\% & 49.66\% \\ 
     & R-BiLSTM-C & 50.24\%  & 50.10\%  & 50.38\% \\ 
     & BiLSTM-Dense & 50.40\%  & 50.80\%  & 50.00\%  \\ 

    \midrule
    \multirow{4}{*}{Instance on secp256k1} 
     & FCN &  49.88\%  & 49.75\% & 50.01\% \\ 
     & R-BiLSTM-C &  50.52\%  & 50.40\% & 50.64\% \\
     & BiLSTM-Dense &  50.14\%  & 49.90\% & 50.38\% \\

     \midrule
     \multirow{4}{*}{Instance on P-256} 
     & FCN &  50.23\%  & 50.12\% & 50.34\% \\ 
     & R-BiLSTM-C&  49.45\%  & 49.60\% & 49.30\% \\
     & BiLSTM-Dense &  49.78\% & 49.50\% & 50.06\% \\
 
    \bottomrule
    \end{tabular}
\end{table}

The results are presented in Tab.~\ref{tab:Steganalysis}, which reveals that even under such a large scale, the detection error rates are still close to 50\%. The false alarm rate and missed detection rate are also consistently near 50\%. These results suggest that it is challenging to distinguish between the distribution of stegotext with secret information and the randomly sampled covertext. 

Through the steganalysis of these three instances, we have demonstrated that our framework can be safely deployed on all commonly used curves, greatly increasing the applicability of public-key steganography. Furthermore, due to the regularity property of admissible encoding, in algorithm Alg. \ref{Alg_SE}, the \textit{Point Hiding} step will terminate the loop in $O(1)$ time for any sampled curve point. In other words, our public key steganography can efficiently cover all curve points. 

Furthermore, in Appendix \ref{Subsec_generalization_to_hyperellipic_curves}, we list a plethora of elliptic curve well-distributed encoding methods that can be applied within our framework to construct public-key steganography and present the algorithm for hyperelliptic curves with genus larger than $1$. Through the tensor exponent function, our scheme can be extended to even work on hyperelliptic curves. 


\section{Discussion and Future Work}

In this paper, we propose a general and complete public-key steganography framework based on admissible encoding which can be employed on all types of curves and utilize all points on the curve. Due to the strict requirements of existing point compression methods on curve parameters, and the inability to establish a surjective mapping from all curve points to a one-dimensional finite field, we consider establishing mappings on two-dimensional or even higher-dimensional finite fields. By utilizing some imperfect but well-distributed encoding techniques through the tensor square (on genus $1$ curves) or tensor exponent (on curves with genus larger than $1$), we construct the powerful tool called admissible encoding. This encoding forms a surjection onto the set of curve points, and under the premise of constructing suitable sampleable inverse functions, it can decode corresponding curve points from multidimensional finite fields. These properties provide theoretical and algorithmic foundations for achieving provably secure public-key steganography.

The significance of the work described in this paper lies in our achievement of extending almost all provably secure steganography methods to the public-key setting while removing all restrictions on curve parameters and allowing encoding for all points. This provides the most universal interface for deploying related higher-level protocols. Subsequent researchers can utilize our work to consider the implementation and deployment of covert protocols on social networks, such as secret sharing or group key agreement, which is a very interesting topic.

Moreover, we aim to broaden the scope of steganography by incorporating additional communication protocols into covert application scenarios \cite{berndt2020universal}. Looking ahead, we envision the potential to create a parallel environment within large-scale applications. In this parallel world, we would have communication tools akin to those in the real world, yet remain undetectable to external observers.

\bibliography{main}

\begin{thebibliography}{10}
\providecommand{\url}[1]{#1}
\csname url@samestyle\endcsname
\providecommand{\newblock}{\relax}
\providecommand{\bibinfo}[2]{#2}
\providecommand{\BIBentrySTDinterwordspacing}{\spaceskip=0pt\relax}
\providecommand{\BIBentryALTinterwordstretchfactor}{4}
\providecommand{\BIBentryALTinterwordspacing}{\spaceskip=\fontdimen2\font plus
\BIBentryALTinterwordstretchfactor\fontdimen3\font minus \fontdimen4\font\relax}
\providecommand{\BIBforeignlanguage}[2]{{%
\expandafter\ifx\csname l@#1\endcsname\relax
\typeout{** WARNING: IEEEtran.bst: No hyphenation pattern has been}%
\typeout{** loaded for the language `#1'. Using the pattern for}%
\typeout{** the default language instead.}%
\else
\language=\csname l@#1\endcsname
\fi
#2}}
\providecommand{\BIBdecl}{\relax}
\BIBdecl

\bibitem{anderson1998limits}
R.~J. Anderson and F.~A. Petitcolas, ``On the limits of steganography,'' \emph{IEEE Journal on selected areas in communications}, vol.~16, no.~4, pp. 474--481, 1998.

\bibitem{marvel1999spread}
L.~M. Marvel, C.~G. Boncelet, and C.~T. Retter, ``Spread spectrum image steganography,'' \emph{IEEE Transactions on image processing}, vol.~8, no.~8, pp. 1075--1083, 1999.

\bibitem{cox2007digital}
I.~Cox, M.~Miller, J.~Bloom, J.~Fridrich, and T.~Kalker, \emph{Digital watermarking and steganography}.\hskip 1em plus 0.5em minus 0.4em\relax Morgan kaufmann, 2007.

\bibitem{simmons1984prisoners}
G.~J. Simmons, ``The prisoners’ problem and the subliminal channel,'' in \emph{Advances in Cryptology: Proceedings of Crypto 83}.\hskip 1em plus 0.5em minus 0.4em\relax Springer, 1984, pp. 51--67.

\bibitem{yang2018rnn}
Z.-L. Yang, X.-Q. Guo, Z.-M. Chen, Y.-F. Huang, and Y.-J. Zhang, ``Rnn-stega: Linguistic steganography based on recurrent neural networks,'' \emph{IEEE Transactions on Information Forensics and Security}, vol.~14, no.~5, pp. 1280--1295, 2018.

\bibitem{fridrich2012rich}
J.~Fridrich and J.~Kodovsky, ``Rich models for steganalysis of digital images,'' \emph{IEEE Transactions on information Forensics and Security}, vol.~7, no.~3, pp. 868--882, 2012.

\bibitem{mielikainen2006lsb}
J.~Mielikainen, ``Lsb matching revisited,'' \emph{IEEE signal processing letters}, vol.~13, no.~5, pp. 285--287, 2006.

\bibitem{zhang2006efficient}
X.~Zhang and S.~Wang, ``Efficient steganographic embedding by exploiting modification direction,'' \emph{IEEE Communications letters}, vol.~10, no.~11, pp. 781--783, 2006.

\bibitem{6949122}
B.~Feng, W.~Lu, and W.~Sun, ``Secure binary image steganography based on minimizing the distortion on the texture,'' \emph{IEEE Transactions on Information Forensics and Security}, vol.~10, no.~2, pp. 243--255, 2015.

\bibitem{pevny2010using}
T.~Pevn{\`y}, T.~Filler, and P.~Bas, ``Using high-dimensional image models to perform highly undetectable steganography,'' in \emph{Information Hiding: 12th International Conference, IH 2010, Calgary, AB, Canada, June 28-30, 2010, Revised Selected Papers 12}.\hskip 1em plus 0.5em minus 0.4em\relax Springer, 2010, pp. 161--177.

\bibitem{ye2017deep}
J.~Ye, J.~Ni, and Y.~Yi, ``Deep learning hierarchical representations for image steganalysis,'' \emph{IEEE Transactions on Information Forensics and Security}, vol.~12, no.~11, pp. 2545--2557, 2017.

\bibitem{boroumand2018deep}
M.~Boroumand, M.~Chen, and J.~Fridrich, ``Deep residual network for steganalysis of digital images,'' \emph{IEEE Transactions on Information Forensics and Security}, vol.~14, no.~5, pp. 1181--1193, 2018.

\bibitem{cachin1998information}
C.~Cachin, ``An information-theoretic model for steganography,'' in \emph{Information Hiding: Second International Workshop, IH’98 Portland, Oregon, USA, April 14--17, 1998 Proceedings}.\hskip 1em plus 0.5em minus 0.4em\relax Springer, 1998, pp. 306--318.

\bibitem{hopper2002provably}
N.~J. Hopper, J.~Langford, and L.~Von~Ahn, ``Provably secure steganography,'' in \emph{Advances in Cryptology—CRYPTO 2002: 22nd Annual International Cryptology Conference Santa Barbara, California, USA, August 18--22, 2002 Proceedings 22}.\hskip 1em plus 0.5em minus 0.4em\relax Springer, 2002, pp. 77--92.

\bibitem{arjovsky2017wasserstein}
M.~Arjovsky, S.~Chintala, and L.~Bottou, ``Wasserstein generative adversarial networks,'' in \emph{International conference on machine learning}.\hskip 1em plus 0.5em minus 0.4em\relax PMLR, 2017, pp. 214--223.

\bibitem{jozefowicz2016exploring}
R.~Jozefowicz, O.~Vinyals, M.~Schuster, N.~Shazeer, and Y.~Wu, ``Exploring the limits of language modeling,'' \emph{arXiv preprint arXiv:1602.02410}, 2016.

\bibitem{prenger2019waveglow}
R.~Prenger, R.~Valle, and B.~Catanzaro, ``Waveglow: A flow-based generative network for speech synthesis,'' in \emph{ICASSP 2019-2019 IEEE International Conference on Acoustics, Speech and Signal Processing (ICASSP)}.\hskip 1em plus 0.5em minus 0.4em\relax IEEE, 2019, pp. 3617--3621.

\bibitem{ziegler2019neural}
Z.~Ziegler, Y.~Deng, and A.~M. Rush, ``Neural linguistic steganography,'' in \emph{Proceedings of the 2019 Conference on Empirical Methods in Natural Language Processing and the 9th International Joint Conference on Natural Language Processing (EMNLP-IJCNLP)}, 2019, pp. 1210--1215.

\bibitem{chen2021distribution}
K.~Chen, H.~Zhou, H.~Zhao, D.~Chen, W.~Zhang, and N.~Yu, ``Distribution-preserving steganography based on text-to-speech generative models,'' \emph{IEEE Transactions on Dependable and Secure Computing}, vol.~19, no.~5, pp. 3343--3356, 2021.

\bibitem{zhang2021provably}
S.~Zhang, Z.~Yang, J.~Yang, and Y.~Huang, ``Provably secure generative linguistic steganography,'' \emph{arXiv preprint arXiv:2106.02011}, 2021.

\bibitem{kaptchuk2021meteor}
G.~Kaptchuk, T.~M. Jois, M.~Green, and A.~D. Rubin, ``Meteor: Cryptographically secure steganography for realistic distributions,'' in \emph{Proceedings of the 2021 ACM SIGSAC Conference on Computer and Communications Security}, 2021, pp. 1529--1548.

\bibitem{deperfectly}
C.~S. de~Witt, S.~Sokota, J.~Z. Kolter, J.~N. Foerster, and M.~Strohmeier, ``Perfectly secure steganography using minimum entropy coupling,'' in \emph{The Eleventh International Conference on Learning Representations}.

\bibitem{dingDiscopProvablySecure2023}
J.~Ding, K.~Chen, Y.~Wang, N.~Zhao, W.~Zhang, and N.~Yu, ``Discop: {{Provably Secure Steganography}} in {{Practice Based}} on ``{{Distribution Copies}}'','' in \emph{2023 {{IEEE Symposium}} on {{Security}} and {{Privacy}} ({{SP}})}, May 2023.

\bibitem{von2004public}
L.~Von~Ahn and N.~J. Hopper, ``Public-key steganography,'' in \emph{Advances in Cryptology-EUROCRYPT 2004: International Conference on the Theory and Applications of Cryptographic Techniques, Interlaken, Switzerland, May 2-6, 2004. Proceedings 23}.\hskip 1em plus 0.5em minus 0.4em\relax Springer, 2004, pp. 323--341.

\bibitem{zhang2024provably}
X.~Zhang, K.~Chen, J.~Ding, Y.~Yang, W.~Zhang, and N.~Yu, ``Provably secure public-key steganography based on elliptic curve cryptography,'' \emph{IEEE Transactions on Information Forensics and Security}, vol.~19, pp. 3148--3163, 2024.

\bibitem{boneh2001identity}
D.~Boneh and M.~Franklin, ``Identity-based encryption from the weil pairing,'' in \emph{Annual international cryptology conference}.\hskip 1em plus 0.5em minus 0.4em\relax Springer, 2001, pp. 213--229.

\bibitem{horwitz2002toward}
J.~Horwitz and B.~Lynn, ``Toward hierarchical identity-based encryption,'' in \emph{International conference on the theory and applications of cryptographic techniques}.\hskip 1em plus 0.5em minus 0.4em\relax Springer, 2002, pp. 466--481.

\bibitem{boneh2003aggregate}
D.~Boneh, C.~Gentry, B.~Lynn, and H.~Shacham, ``Aggregate and verifiably encrypted signatures from bilinear maps,'' in \emph{Advances in Cryptology—EUROCRYPT 2003: International Conference on the Theory and Applications of Cryptographic Techniques, Warsaw, Poland, May 4--8, 2003 Proceedings 22}.\hskip 1em plus 0.5em minus 0.4em\relax Springer, 2003, pp. 416--432.

\bibitem{icart2009hash}
T.~Icart, ``How to hash into elliptic curves,'' in \emph{Annual International Cryptology Conference}.\hskip 1em plus 0.5em minus 0.4em\relax Springer, 2009, pp. 303--316.

\bibitem{farashahi2013indifferentiable}
R.~R. Farashahi, P.-A. Fouque, I.~Shparlinski, M.~Tibouchi, and J.~Voloch, ``Indifferentiable deterministic hashing to elliptic and hyperelliptic curves,'' \emph{Mathematics of Computation}, vol.~82, no. 281, pp. 491--512, 2013.

\bibitem{fouque2012indifferentiable}
P.-A. Fouque and M.~Tibouchi, ``Indifferentiable hashing to barreto--naehrig curves,'' in \emph{Progress in Cryptology--LATINCRYPT 2012: 2nd International Conference on Cryptology and Information Security in Latin America, Santiago, Chile, October 7-10, 2012. Proceedings 2}.\hskip 1em plus 0.5em minus 0.4em\relax Springer, 2012, pp. 1--17.

\bibitem{brier2010efficient}
E.~Brier, J.-S. Coron, T.~Icart, D.~Madore, H.~Randriam, and M.~Tibouchi, ``Efficient indifferentiable hashing into ordinary elliptic curves,'' in \emph{Advances in Cryptology--CRYPTO 2010: 30th Annual Cryptology Conference, Santa Barbara, CA, USA, August 15-19, 2010. Proceedings 30}.\hskip 1em plus 0.5em minus 0.4em\relax Springer, 2010, pp. 237--254.

\bibitem{katzenbeisser2002defining}
S.~Katzenbeisser and F.~A. Petitcolas, ``Defining security in steganographic systems,'' in \emph{Security and Watermarking of Multimedia Contents IV}, vol. 4675.\hskip 1em plus 0.5em minus 0.4em\relax SPIE, 2002, pp. 50--56.

\bibitem{yang2019provably}
K.~Yang, K.~Chen, W.~Zhang, and N.~Yu, ``Provably secure generative steganography based on autoregressive model,'' in \emph{Digital Forensics and Watermarking: 17th International Workshop, IWDW 2018, Jeju Island, Korea, October 22-24, 2018, Proceedings}.\hskip 1em plus 0.5em minus 0.4em\relax Springer, 2019, pp. 55--68.

\bibitem{hopper2005steganographic}
N.~Hopper, ``On steganographic chosen covertext security,'' in \emph{Automata, Languages and Programming: 32nd International Colloquium, ICALP 2005, Lisbon, Portugal, July 11-15, 2005. Proceedings 32}.\hskip 1em plus 0.5em minus 0.4em\relax Springer, 2005, pp. 311--323.

\bibitem{shallue2006construction}
A.~Shallue and C.~E. van~de Woestijne, ``Construction of rational points on elliptic curves over finite fields,'' in \emph{Algorithmic Number Theory: 7th International Symposium, ANTS-VII, Berlin, Germany, July 23-28, 2006. Proceedings 7}.\hskip 1em plus 0.5em minus 0.4em\relax Springer, 2006, pp. 510--524.

\bibitem{ulas2007rational}
M.~Ulas, ``Rational points on certain hyperelliptic curves over finite fields,'' \emph{arXiv preprint arXiv:0706.1448}, 2007.

\bibitem{shoup1995new}
V.~Shoup, ``A new polynomial factorization algorithm and its implementation,'' \emph{Journal of Symbolic Computation}, vol.~20, no.~4, pp. 363--397, 1995.

\bibitem{barreto2005pairing}
P.~S. Barreto and M.~Naehrig, ``Pairing-friendly elliptic curves of prime order,'' in \emph{International workshop on selected areas in cryptography}.\hskip 1em plus 0.5em minus 0.4em\relax Springer, 2005, pp. 319--331.

\bibitem{yang2019fast}
Z.~Yang, Y.~Huang, and Y.-J. Zhang, ``A fast and efficient text steganalysis method,'' \emph{IEEE Signal Processing Letters}, vol.~26, no.~4, pp. 627--631, 2019.

\bibitem{niuHybridRBILSTMCNeural2019}
Y.~Niu, J.~Wen, P.~Zhong, and Y.~Xue, ``A {{Hybrid R-BILSTM-C Neural Network Based Text Steganalysis}},'' \emph{IEEE Signal Processing Letters}, vol.~26, no.~12, pp. 1907--1911, Dec. 2019.

\bibitem{yangLinguisticSteganalysisDensely2020}
H.~Yang, Y.~Bao, Z.~Yang, S.~Liu, Y.~Huang, and S.~Jiao, ``Linguistic {{Steganalysis}} via {{Densely Connected LSTM}} with {{Feature Pyramid}},'' in \emph{Proceedings of the 2020 {{ACM Workshop}} on {{Information Hiding}} and {{Multimedia Security}}}.\hskip 1em plus 0.5em minus 0.4em\relax {Denver CO USA}: {ACM}, Jun. 2020, pp. 5--10.

\bibitem{berndt2020universal}
S.~Berndt and M.~Li{\'s}kiewicz, ``On the universal steganography of optimal rate,'' \emph{Information and Computation}, vol. 275, p. 104632, 2020.

\bibitem{niederreiter2009algebraic}
H.~Niederreiter and C.~Xing, \emph{Algebraic geometry in coding theory and cryptography}.\hskip 1em plus 0.5em minus 0.4em\relax Princeton University Press, 2009.

\bibitem{skalba2005points}
M.~Ska{\l}ba, ``Points on elliptic curves over finite fields,'' \emph{Acta Arithmetica}, vol. 117, no.~3, pp. 293--301, 2005.

\bibitem{fouque2010deterministic}
P.-A. Fouque and M.~Tibouchi, ``Deterministic encoding and hashing to odd hyperelliptic curves,'' in \emph{International Conference on Pairing-Based Cryptography}.\hskip 1em plus 0.5em minus 0.4em\relax Springer, 2010, pp. 265--277.

\bibitem{farashahi2011hashing}
R.~R. Farashahi, ``Hashing into hessian curves,'' in \emph{International Conference on Cryptology in Africa}.\hskip 1em plus 0.5em minus 0.4em\relax Springer, 2011, pp. 278--289.

\bibitem{kammerer2010encoding}
J.-G. Kammerer, R.~Lercier, and G.~Renault, ``Encoding points on hyperelliptic curves over finite fields in deterministic polynomial time,'' in \emph{Pairing-Based Cryptography-Pairing 2010: 4th International Conference, Yamanaka Hot Spring, Japan, December 2010. Proceedings 4}.\hskip 1em plus 0.5em minus 0.4em\relax Springer, 2010, pp. 278--297.

\end{thebibliography}
\bibliographystyle{IEEEtran}

\section{Appendix}\label{app: Appendix}

\subsection{Generalization to Hyperelliptic Curves}\label{Subsec_generalization_to_hyperellipic_curves}

The previous public-key steganography encoder (SE), as defined in Algorithm \ref{Alg_SE}, and the public-key steganography decoder (SD), as defined in Algorithm \ref{Alg_SD}, are specifically designed for deployment on curves of genus 1. To adapt these algorithms for use on hyperelliptic curves, which have a higher genus, it is necessary to transition from employing the tensor square function to utilizing the tensor exponent function.

Let the curve $E(\mathbb{F}_p)$ of genus $d$ defined on finite field $\mathbb{F}_p$ and $f$ is its computable $B$-well-distributed encoding. Other definitions are similar to those mentioned previously. We list below the provably secure public-key steganography encoder (SE, Alg.\ref{Alg_SE_hyper}) and decoder (SD, Alg.\ref{Alg_SD_hyper}) algorithms that can be deployed on hyperelliptic curves.

According to Lemma \ref{lem_well_distributed_encoding_bound}, we can prove the admissibility of the tensor exponent function in a manner similar to Lemma \ref{lem_Admissibility_of_tensor_square_function}. With the remaining parts unchanged, our public-key steganography scheme can withstand CHA.


\begin{algorithm}[htp]
	\renewcommand{\algorithmicrequire}{\textbf{INPUT:}}
	\renewcommand{\algorithmicensure}{\textbf{OUTPUT:}}
	\caption{\textbf{Public-key Steganography Encoder (SE) Extended to Hyperelliptic Curves}
	                }
	\label{Alg_SE_hyper}
	\begin{algorithmic}[1]
	    \REQUIRE {$\mathcal{E},\ m,\ (p,\  E\left(\mathbb{F}_{p}\right),\ g,\ q,\ PK),\ (f,\ \mathcal{I})$}
	    \ENSURE {$s_1,s_2,\cdots,s_*$}
            \STATE \#\# \textit{Key Deriving}
		\STATE Pick $a \in [0,\ n-1]$ at random;
		\STATE $P = a \cdot g$
            \STATE $K = H(a \cdot PK)$
            \STATE \#\# \textit{Point Hiding}
            \WHILE{True}
		\STATE Pick $v_1\  \cdots\  v_d \in \mathbb{F}_{p}$ at random;
		\STATE $D \leftarrow \mathcal{I}(P-f(v_1)-f(v_2)-\cdots-f(v_d))$;
            \STATE pick $i$ uniformly at random in \#$D$;
            \STATE $v_{d+1} \leftarrow$ $i$-th element of $D$;
            \STATE \textbf{if} $u=\varnothing$ \textbf{continue};
            \STATE \ \ \ \ \ \ \textbf{else} \textbf{break};
		\ENDWHILE
            \STATE \#\# \textit{Bias Eliminating}
            \STATE Choose $r$-bit redundancy:
            \STATE Pick $r_1,r_2,\cdots,r_{d+1} \in\left\{0, \ldots,\left\lfloor\frac{2^{k+r}-v_i}{q}\right\rfloor\right\}$ 
            \STATE $\Tilde{v}_i = v_i+r_1p,\ \forall\ i \in \{1\cdots d+1\}$
            \STATE \#\# \textit{Final Encoding}
            \STATE $C_1 = \Tilde{v_1} || \Tilde{v_2} || \cdots || \Tilde{v_{d+1}}$
            \STATE $C_2 = E_K(K,m)$
            \STATE $s_1,\ s_2,\cdots,\ s_* = \mathcal{E}(C_1 || C_2)$
	\end{algorithmic}  
\end{algorithm}
The algorithm is largely similar to the one described above, with the sole distinction being in the Point Hiding part, where we utilize a tensor square construction of admissible encoding with \(s=d+1\). As a result, the first \(d\) finite field elements are generated by random sampling, and the last element is obtained through sampling using the sampleable inverse function of \(f\). Hence, the length of \(C_1\) is a bit sequence of \(d+1\) segments.

We have listed a plethora of elliptic curve well-distributed encoding method in Tab. \ref{tab_known_encodings} that can be applied within our framework to construct public-key steganography. Through the tensor exponent function, our scheme can be extended to even work on hyperelliptic curves.

\begin{algorithm}[htp]
	\renewcommand{\algorithmicrequire}{\textbf{INPUT:}}
	\renewcommand{\algorithmicensure}{\textbf{OUTPUT:}}
	\caption{\textbf{Public-key Steganography Decoder (SD) Extended to Hyperelliptic Curves}
	                }
	\label{Alg_SD_hyper}
	\begin{algorithmic}[1]
	    \REQUIRE {$\mathcal{D},\ s_{1..*},\ (p,\  E\left(\mathbb{F}_{p}\right),\ g,\ n,\ SK),\ f$}
	    \ENSURE {$m$}
            \STATE Split $C = \mathcal{D}(s_{1..*})$ into $C_1,\ C_2$;
            \STATE $(\Tilde{v}_1,\ \Tilde{v}_2,\ \cdots,\ \Tilde{v}_{d+1})=C_1$
            \STATE $v_i = \Tilde{v_i}\ mod\ p,\ \forall\ i \in \{1,\cdots,d+1\}$
            \STATE $P = f(v_1) + f(v_2) + \cdots +f(v_{d+1})$
            \STATE $K = H(SK \cdot P)$
            \STATE $m = D_K(K, C_2)$
	\end{algorithmic}  
\end{algorithm}

\subsection{List of Key Lemmas}
\begin{lemma}\label{lem_Hasse_bound}\textbf{(Hasse Bound)} \cite{niederreiter2009algebraic}
    For curve $E\left(\mathbb{F}_{p}\right)$ of genus $1$ defined over finite field $\mathbb{F}_{p}$, we have:
\begin{equation}
    |\#E\left(\mathbb{F}_{p}\right)-p-1| \leq 2 \sqrt{p}
\end{equation}
\end{lemma}

\begin{lemma}\label{lem_well_distributed_encoding_bound}
\textbf{\textup{(\cite{farashahi2013indifferentiable}, Corollary 4)}}
If $f: \mathbb{F}_{p} \rightarrow E\left(\mathbb{F}_{p}\right)$ is a $B$-well-distributed encoding into a curve $E\left(\mathbb{F}_{p}\right)$, then the statistical distance between the distribution defined by a tensor exponent function $f^{\otimes s}$ on $E\left(\mathbb{F}_{p}\right)$ and the uniform distribution is bounded as:
\begin{equation}
    \sum_{D \in E\left(\mathbb{F}_{p}\right)}\left|\frac{N_{s}(D)}{p^{s}}-\frac{1}{\# E\left(\mathbb{F}_{p}\right)}\right| \leq \frac{B^{s}}{p^{s / 2}} \sqrt{\# E\left(\mathbb{F}_{p}\right)},
\end{equation}

\end{lemma}

\begin{lemma}\label{lem_Icart_bound}
\textbf{\textup{(\cite{farashahi2013indifferentiable}, Theorem 8)}}
Let $f$ be Icart's encoding function (\ref{equation_Icart_encoding}). For any nontrivial character $\chi$ of $E\left(\mathbb{F}_{p}\right)$, the character sum
$S_f(\chi)$ given by (\ref{equation_S_f}) satisfies:
\begin{equation}
    \left|S_{f}(\chi)\right| \leq 12 \sqrt{p}+3.
\end{equation}

In other words, Icart's encoding is a ($12+3p^{-\frac{1}{2}}$)-well-distributed encoding.

\end{lemma}

\begin{lemma}\label{lem_SW_bound}
\textbf{\textup{(\cite{fouque2012indifferentiable}, Section 5)}}
Let $f$ be SW encoding function (\ref{equation_SW_encoding}). For any nontrivial character $\chi$ of $E\left(\mathbb{F}_{p}\right)$, the character sum
$S_f(\chi)$ given by (\ref{equation_S_f}) satisfies:
\begin{equation}
    \left|S_{f}(\chi)\right| \leq 62 \sqrt{p}+O(1).
\end{equation}

In other words, Icart's encoding is a ($62+O(p^{-\frac{1}{2}})$)-well-distributed encoding.

\end{lemma}

\begin{lemma}\label{lem_SWu_bound}
\textbf{\textup{(\cite{farashahi2013indifferentiable}, Theorem 15)}}
Let $f$ be the SWU encoding function (\ref{equation_SWU_encoding}). For any nontrivial character $\chi$ of $E\left(\mathbb{F}_{p}\right)$, the character sum
$S_f(\chi)$ given by (\ref{equation_S_f}) satisfies:
\begin{equation}
    \left|S_{f}(\chi)\right| \leq 52 \sqrt{p}+151.
\end{equation}

In other words, SWU encoding is a $(52+151p^{-\frac{1}{2}})$-well-distributed encoding.

\end{lemma}

\begin{table*}[t]
\renewcommand{\arraystretch}{1.5}
\caption{Known deterministic well-distributed encodings to commonly used elliptic curves and hyperelliptic curves.}
\label{tab_known_encodings}
\resizebox{\textwidth}{!}{
\begin{tabular}{|c|c|c|c|c|}
\hline
\textbf{char.} & \textbf{curve equation} & \textbf{genus} & \textbf{encoding} & \textbf{conditions on \( p \)} \\
\hline
\(\neq 2, 3\) & \( y^2 = x^3 + ax + b \) & 1 & Skalba\cite{skalba2005points} & -- \\
\hline
\(\neq 2, 3\) & \( y^2 = x^3 + ax + b \) & 1 & SWU\cite{brier2010efficient} & \( p \equiv 3 \mod 4 \) \\
\hline
\( \neq 2, 3 \) & \( y^2 = x^{2g+1} + ax + b \) & \( g \) & Ulas\cite{ulas2007rational} & -- \\
\hline
\( \neq 2, 3 \) & \( y^2 = x^{2g+1} + a_1x^{2g-1} + \cdots + a_gx \) & \( g \) & FT\cite{fouque2010deterministic} & \( p \equiv 3 \mod 4 \) \\
\hline
any & any & 1 & SW\cite{shallue2006construction} & --\\
\hline
3 & \( y^2 = x^3 + ax^2 + b \) & 1 & Brier \textit{et al.}\cite{brier2010efficient} & -- \\
\hline
\( \neq 2, 3 \) & \( y^2 = x^3 + ax + b \) & 1 & Icart\cite{icart2009hash} & \( p = 2 \mod 3 \) \\
\hline
\( \neq 2, 3 \) & \( x^3 + y^3 + 1 = 3dxy \) & 1 & F\cite{farashahi2011hashing} \& KLR\cite{kammerer2010encoding} & \( p = 2 \mod 3 \) \\
\hline
\( \neq 2, 3 \) & \( x^3 + (y + c)(3x + 2a + 2b/y) = 0 \) & 2 & KLR\cite{kammerer2010encoding} & \( p \equiv 2 \mod 3 \) \\
\hline
\( \neq 2, 3 \) & \( y^2 = x^{2d} + x^d + a \) & \( d-1 \) & KLR\cite{kammerer2010encoding} & \( (d, p - 1) = 1 \) \\
\hline
\( \neq 2, 3 \) & \( y^2 = p_{a,b}^{(d)}(x) \) & \( \frac{d-1}{2} \) & KLR\cite{kammerer2010encoding} & \( p \equiv 2 \mod 3,(d, p - 1) = 1  \) \\
\hline
2 & \( y^2 + y = p_{a,b}^{(d)}(x) \) & \( \frac{d-1}{2} \) & KLR\cite{kammerer2010encoding} & \( p = 2 \mod 3 \) \\
\hline
2 & \( y^2 + xy = x^3 + ax^2 + b \) & 1 & Icart\cite{icart2009hash} & \( p = 2 \mod 3 \) \\
\hline
\end{tabular}
}
\end{table*}
\begin{IEEEbiography}[{\includegraphics[width=1in,height=1.25in,clip,keepaspectratio]{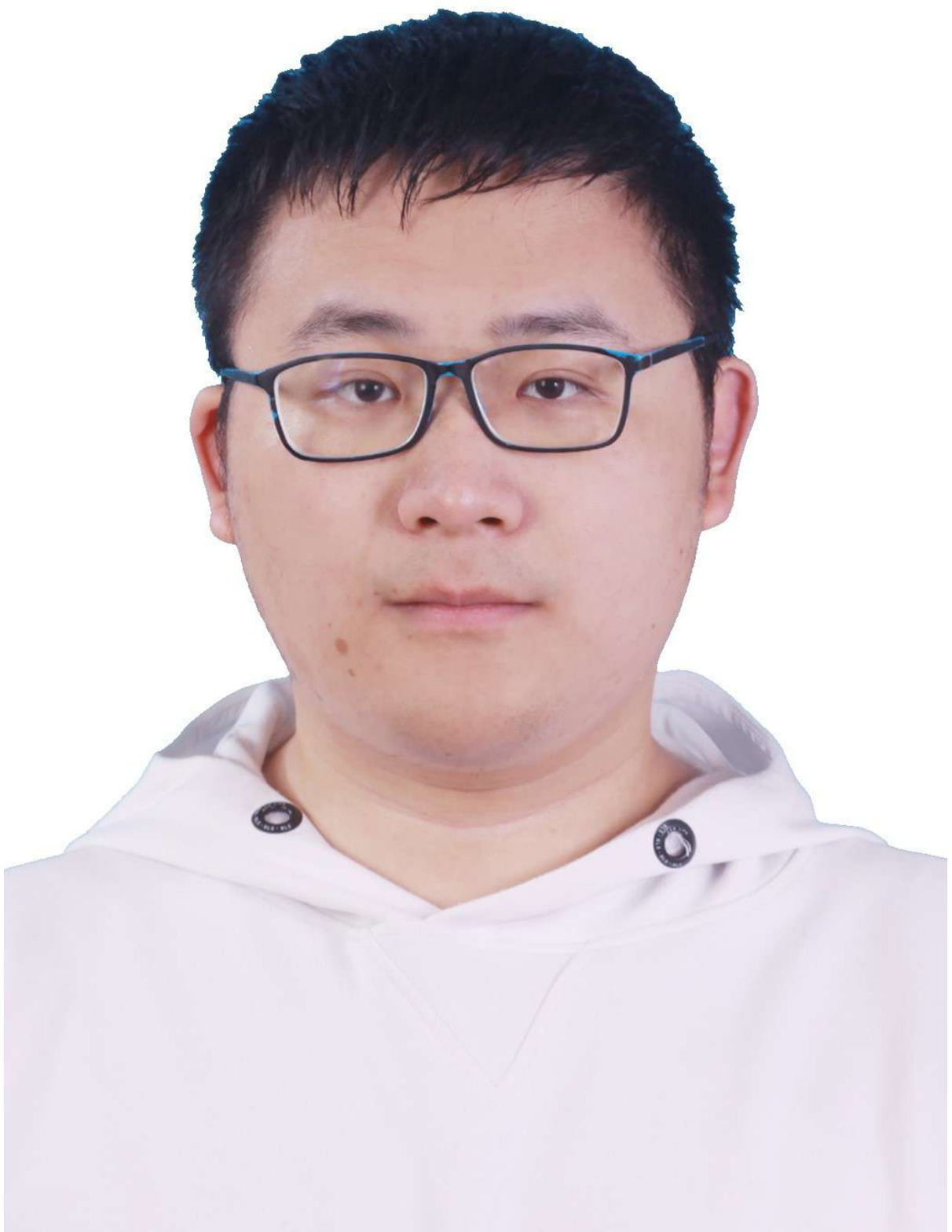}}]{Xin Zhang}
    received his B.S. degree in 2022 from University of Science and Technology of China (USTC). Currently, he is pursuing a master's degree in engineering at USTC. His research interests include information hiding, applied cryptography, and deep learning.
\end{IEEEbiography}

\begin{IEEEbiography}[{\includegraphics[width=1in,height=1.25in,clip,keepaspectratio]{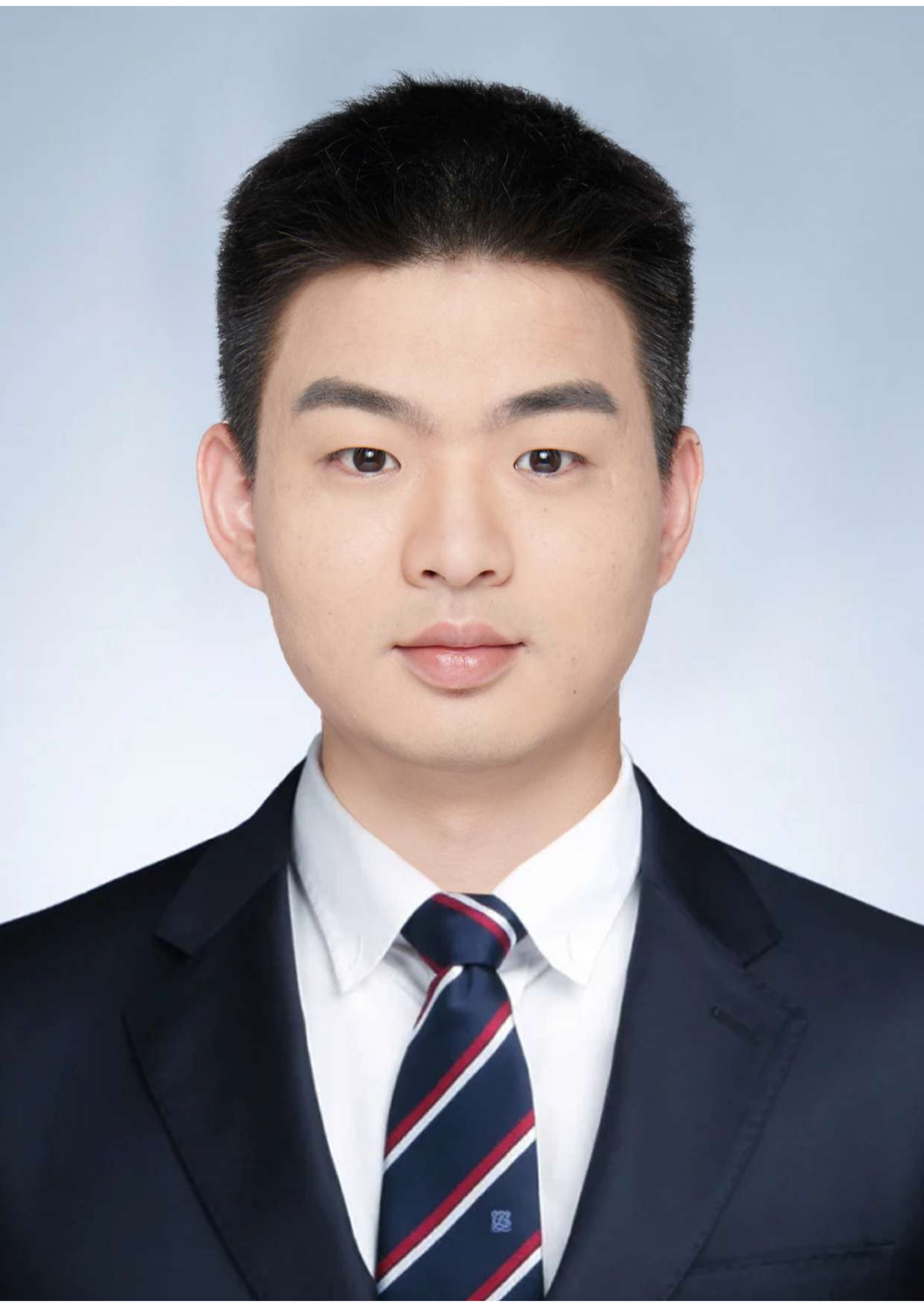}}]{Kejiang Chen}
    received his B.ENG. degree in 2015 from Shanghai University (SHU) and a Ph.D. degree in 2020 from the University of Science and Technology of China (USTC). Currently, he is an associate research fellow at the University of Science and Technology of China. His research interests include information hiding, image processing, and deep learning.
\end{IEEEbiography}

\begin{IEEEbiography}[{\includegraphics[width=1in,height=1.25in,clip,keepaspectratio]{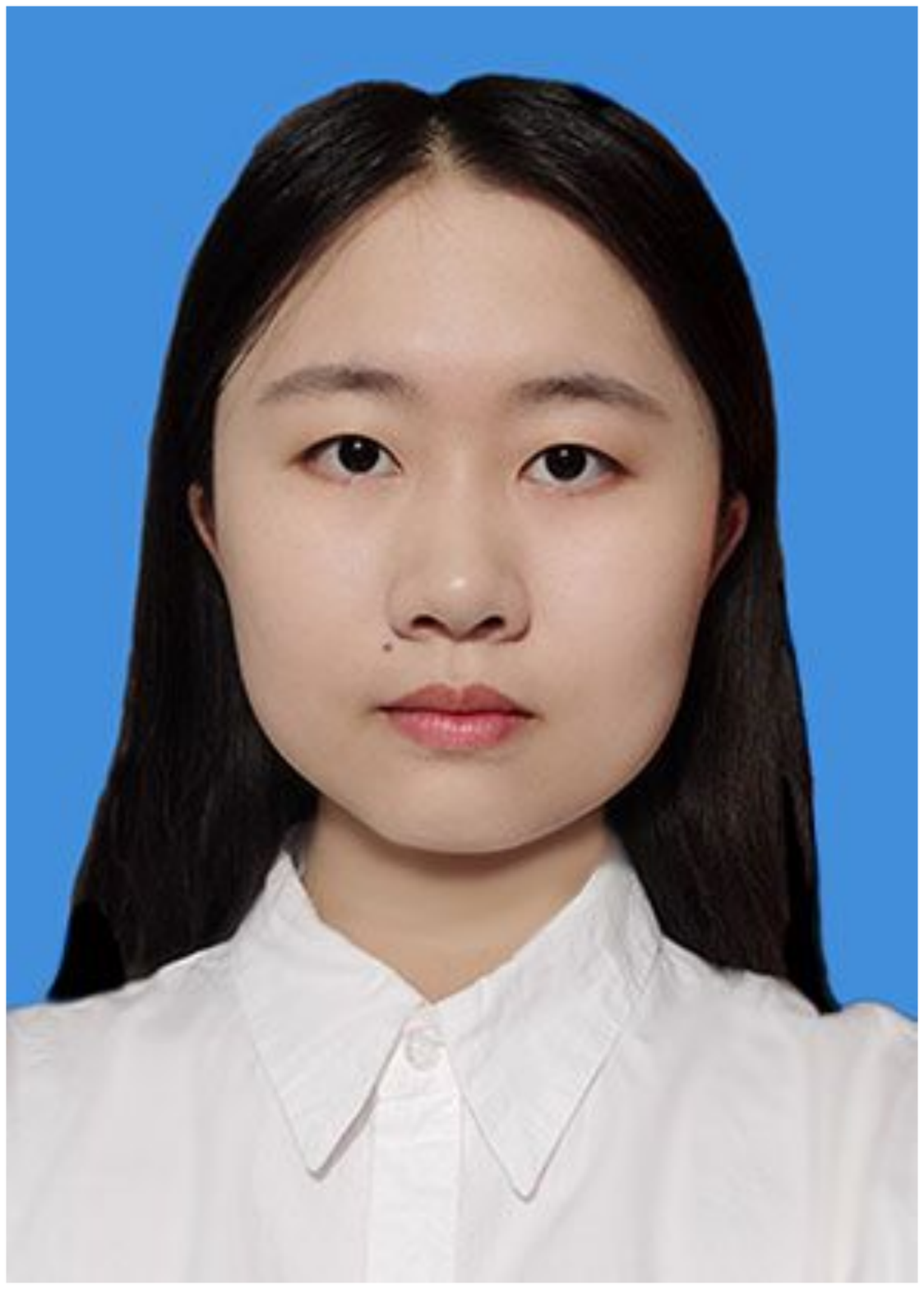}}]{Na Zhao}
    received her B.S. degree in 2017 from Zhengzhou University (ZZU). Currently, she is working toward the Ph.D. degree in the University of Science and Technology of China. Her research interests include information hiding, and deep learning security.
\end{IEEEbiography}

\begin{IEEEbiography}
[{\includegraphics[width=1in,height=1.25in,clip,keepaspectratio]{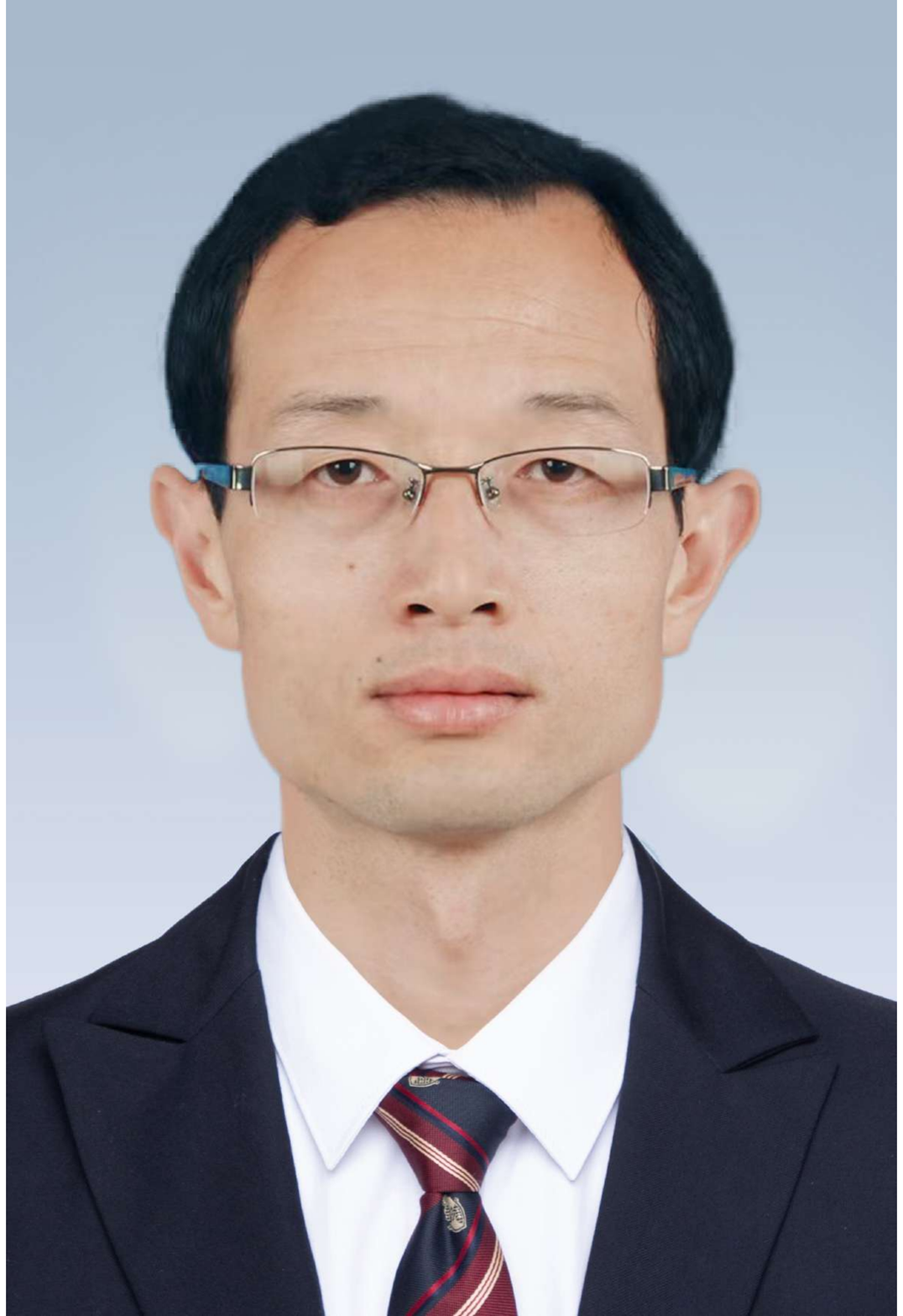}}]{Weiming Zhang}
received his M.S. degree and Ph.D. degree in 2002 and 2005, respectively, from the Zhengzhou Information Science and Technology Institute, P.R. China. Currently, he is a professor at the School of Information Science and Technology, University of Science and Technology of China. His research interests include information hiding and multimedia security.
\end{IEEEbiography}

\begin{IEEEbiography}[{\includegraphics[width=1in,height=1.25in,clip,keepaspectratio]{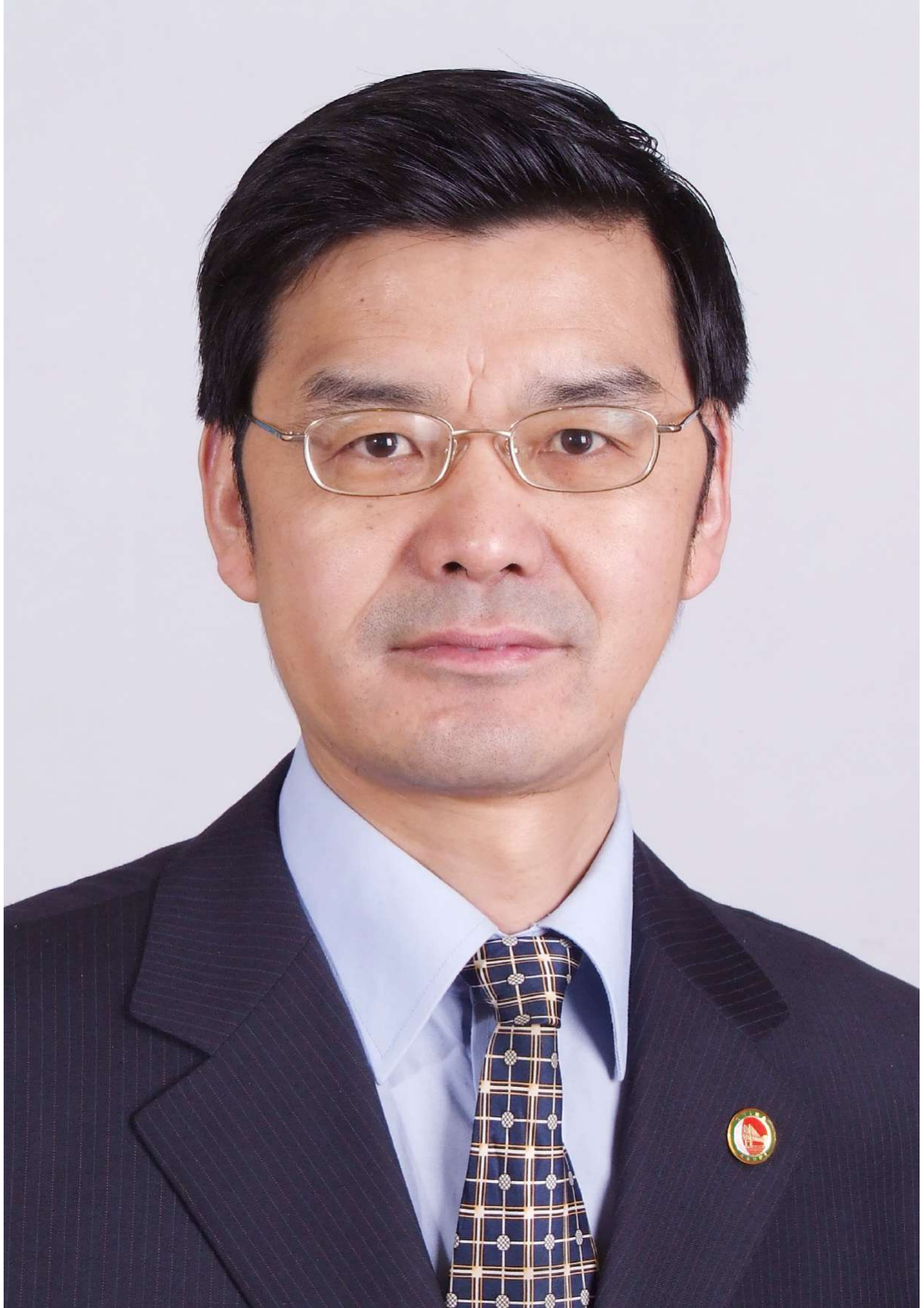}}]{Nenghai Yu}
received his B.S. degree in 1987 from Nanjing University of Posts and Telecommunications, an M.E. degree in 1992 from Tsinghua University, and a Ph.D. degree in 2004 from the University of Science and Technology of China, where he is currently a professor. His research interests include multimedia security, multimedia information retrieval, video processing, and information hiding.
\end{IEEEbiography}

\end{document}